\documentclass[%
 reprint,
superscriptaddress,
amsmath,amssymb, 
aps,
]{revtex4-2}

\usepackage[utf8]{inputenc}
\usepackage[english]{babel}
\usepackage{dsfont}
\usepackage{ mathrsfs }
\usepackage{amssymb}
\usepackage{graphicx}%
\usepackage{chemarr}
\usepackage{dcolumn}
\usepackage{bm}
\usepackage{amsmath}
\usepackage{varioref}
\usepackage{comment}
\usepackage[bottom]{footmisc}
\usepackage{xcolor}

\usepackage{physics}
\usepackage{listings}

\newcommand{\RN}[1]{%
  \textup{\uppercase\expandafter{\romannumeral#1}}%
}

\usepackage{graphicx}

\usepackage{floatrow}
\usepackage{graphicx}
\usepackage{booktabs}
\usepackage{hyperref}

\usepackage[labelformat=simple,subrefformat=parens,labelformat=parens, font=footnotesize]{subfig}
\usepackage{lipsum}
\usepackage{caption}
\begin{document}


\title{Theory for sequence selection via phase separation and oligomerization 
}

\author{Ivar S. Haugerud}
\affiliation{
 Faculty of Mathematics, Natural Sciences, and Materials Engineering: Institute of Physics, University of Augsburg, Universit\"atsstra\ss e~1, 86159 Augsburg, Germany
}

\author{Giacomo Bartolucci}
\affiliation{
 Faculty of Mathematics, Natural Sciences, and Materials Engineering: Institute of Physics, University of Augsburg, Universit\"atsstra\ss e~1, 86159 Augsburg, Germany
}
\affiliation{Department of Physics Universitat de Barcelona, Carrer de Martí i Franquès 1-11, 08028 Barcelona, Spain}

\author{Dieter Braun}

\affiliation{Systems Biophysics, Ludwig Maximilian University Munich, 80799 Munich, Germany}

\author{Christoph A. Weber}

\affiliation{
 Faculty of Mathematics, Natural Sciences, and Materials Engineering: Institute of Physics, University of Augsburg, Universit\"atsstra\ss e~1, 86159 Augsburg, Germany
}

\date{\today}

\begin{abstract}
Non-equilibrium selection pressures were proposed for the formation of oligonucleotides with rich functionalities encoded in their sequences, such as catalysis. 
Since phase separation was shown to direct various chemical processes, we ask whether condensed phases can provide mechanisms for sequence selection. 
To answer this question, we use non-equilibrium thermodynamics and describe the reversible oligomerization of different monomers to sequences at non-dilute conditions prone to phase separation.
We find that when sequences oligomerize, their interactions give rise to phase separation, boosting specific sequences' enrichment and depletion.
Our key result is that phase separation gives rise to a selection pressure for the oligomerization of specific sequence patterns when fragmentation maintains the system away from equilibrium.
Specifically, 
slow fragmentation favors alternating sequences that interact well with their environment (more cooperative), while fast fragmentation selects sequences with extended motifs capable of specific sequence interactions (less cooperative).
Our results highlight that out-of-equilibrium condensed phases  could provide versatile hubs for Darwinian-like evolution toward functional sequences, both relevant for the molecular origin of life and de novo life.
\end{abstract}
\maketitle

\section{Introduction}

\begin{figure*}
    \centering
    \makebox[\textwidth][c]{
    {\includegraphics[width=0.95\textwidth]{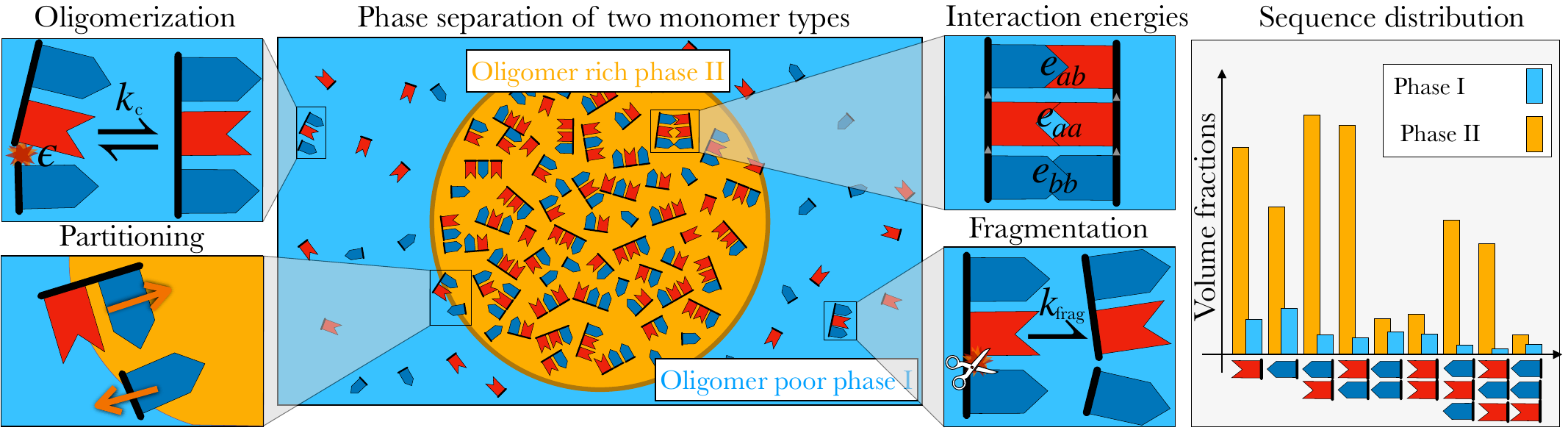}}}
     \caption{
    \textbf{
    Different monomers oligomerize to sequences that can fragment and interact giving rise to oligomer-rich and oligomer-poor  phases:} 
    Two distinct monomeric units $a$ and $b$ reversibly oligomerize with a rate coefficient $k_c$ and  a free energy $\epsilon$ associated with each bond along an oligomer.
     Sequence-dependent interactions among oligomers can lead to  oligomer-rich and oligomer-poor phases mediating the partitioning of monomers and sequences. 
    Sequences interact through monomer-monomer interactions, with interaction parameters $e_{ab}$, $e_{aa}$ and $e_{bb}$.
    We also discuss the case when the system is maintained away from equilibrium by a fragmentation pathway occurring with a rate coefficient $k_\text{frag}$.
Due to non-dilute conditions, phase separation and reversible oligomerization are mutually coupled. This coupling gives rise to complex sequence distributions that differ between the two coexisting phases, I and II.}
    \label{fig:1}
\end{figure*}

Their ability to store genetic information and catalyze chemical reactions makes oligonucleotides, such as DNA and RNA, key building blocks for the molecular origin of life~\cite{robertson_origins_2012,kruger_self-splicing_1982,fedor_catalytic_2005,gilbert_origin_1986}.
Without any specialized biological molecules, such as protein-based enzymes on early Earth, the mechanisms  that can give rise to the emergence of long oligomers with prebiotic functions remain elusive~\cite{birikh_structure_1997,deck_efficient_2011,mariani_light-releasable_2018,walton_mechanism_2019,wunnava_acid-catalyzed_2021}. 
A particularly aspired function is folding into a secondary configuration capable of replicating unfolded oligonucleotide sequences. 
However, a fundamental problem is the exponentially large number of different sequences, particularly for the minimal sequence length required for complex functionalities such as replication~\cite{mccall1992minimal}. 
This problem renders the need for selection mechanisms at the molecular origin of life in the early stages when sequences grow in length and start exploring their exponential sequence space.
Various selection mechanisms were proposed based on the non-equilibrium conditions presumably acting on early Earth~\cite{ianeselli2023physical}, such as  
biased replication~\cite{tkachenko_spontaneous_2015,tkachenko_onset_2018}, 
salt and temperature gradients~\cite{mast_escalation_2013,matreux_heat_2024}, wet-dry~\cite{fares2020impact, maguire_physicochemical_2021,haugerud_nonequilibrium_2024} or freeze-thaw cycles~\cite{mutschler_freezethaw_2015}, shear-driven replication~\cite{eleveld_competitive_2024,liu_light-driven_2024}, accumulation at liquid-vapor interfaces~\cite{morasch_heated_2019} or mineral surfaces~\cite{mizuuchi_mineral_2019}, template aided-ligation~\cite{kudella_structured_2021,calacaserrao_replication_2024,calaca_serrao_high-fidelity_2024}, and finally via phase separated condensates~\cite{bartolucci_sequence_2023}.

The significance of phase-separated condensates in providing a selection mechanism at the molecular origin of life has been suggested roughly a century ago~\cite{haldane1929origin,oparin_origin_1938, fox1976evolutionary, tirard2017jbs}.
These ideas have been recently revisited due to advances in sequencing technologies capable of resolving large sequence distributions~\cite{hu2021next}, complemented by observations of phase separation of oligonucleotide sequences under prebiotic conditions. 
In particular, it was shown that the attractive interactions among different oligomeric sequences can result in the formation of additional phases such as hydrogels~\cite{sato_sequence-based_2020,xing_microrheology_2018}, 
liquid crystalline phases~\cite{nakata_end--end_2007,zanchetta_liquid_2008},
and coacervates~\cite{aumiller_rna-based_2016,jeon_salt-dependent_2018,bartolucci_sequence_2023,bartolucci_interplay_2024}. 
Interestingly, the saturation concentrations above which oligonucleotides such as DNA and RNA phase separate can be rather low, in the order of $\mu M$, due to the strong interaction strength ($5k_BT$) of complementary basis pairs~\cite{aumiller_rna-based_2016,mitrea_phase_2016}. Such low saturation concentrations make  oligonucleotide phase separation a likely and robust scenario, even under varying physicochemical conditions at early Earth.
Furthermore, non-equilibrium thermodynamics implies that the chemical processes such as polymerization, ligation, and fragmentation  of oligonucleotides are generically coupled with phase separation of oligonucleotides~\cite{bauermann_chemical_2022, zwicker2022intertwined, bartolucci_interplay_2024}. This coupling gives rise to feedback between the evolving sequence distribution and their propensity to form condensed phases. 
In other words, when considering the condensed phases  as phenotypes and their local sequence distributions as genotypes, thermodynamics provides a coupling between phenotype and genotype -- a prerequisite for Darwinian evolution.

The coupling between chemical processes and phase separation is expected to mediate a selection mechanism that depends on the inter-sequence interactions.
Indeed, phase separation was shown to gradually enrich 
specific sequences from a pool composed of a few designed sequences~\cite{bartolucci_sequence_2023}.
However, 
it remains an open question whether the interaction among sequences and particularly phase separation, 
can  
direct the oligomerization of sequences.
If phase separation can efficiently direct oligomerization, it could  represent a physicochemical mechanism that reduces the  occupations of the exponentially growing sequence space.   
The coupling between sequence distributions and phase separation becomes especially interesting in the presence of competing effects to aggregation, namely hydrolysis~\cite{goppel_thermodynamic_2022,calaca_serrao_high-fidelity_2024}. Phase separation crates a phase both oligomer-rich and solvent-poor~\cite{bartolucci_interplay_2024}, possibly serving as a protective environment against hydrolysis steering the fragmentation of specific sequences.

In this work, we developed a theoretical framework using non-equilibrium thermodynamics to describe the oligomerization  of sequences at non-dilute conditions. 
Such conditions require accounting for the interactions among all sequences, including the possibility of phase coexistence, as illustrated in Fig.~\ref{fig:1}. Using our theoretical framework, we find that the interactions among sequences direct oligomerization toward sequences that phase separate from the solvent and form a condensed phase. 
We show that phase separation favors forming more cooperatively interacting sequences over specialized sequences that interact only weakly with most other sequences. 
To unravel the role of phase separation in sequence selection, we maintain the system out-of-equilibrium by fragmentation, inspired by the hydrolysis of oligonucleotides.  
We find that  slow fragmentation rates select for sequences of large information content, while low-information sequences with highly correlated sequence motifs emerge for large enough fragmentation rates.
Thus, phase separation mediates a selection pressure for sequences through selective oligomerization. 
Our results indicate that condensed phases can act as spatially localized phenotypes for Darwinian-like evolution toward functional sequences. 

\section{Theory for oligomerization at phase equilibrium}

To describe the dynamics of sequences in phase-separated systems, we use the volume fraction, $\phi_i^\alpha$, where $\alpha$ denotes the phase. 
Here, the component label $i$ includes monomers and oligomers summarised in the set $\sigma$. For two monomeric building blocks, denoted $a$ and $b$, the union of the set of monomers $\sigma_m=\{a,b\}$ and sequences $\sigma_o=\{aa,ab,ba,bb,\ldots\}$ gives all non-solvent components $\sigma=\sigma_m\cup\sigma_o$. 
The volume fraction of a component, $\phi_i^\alpha=\nu_iN_i^\alpha/V^\alpha$, quantifies the volume occupied by $N_i^\alpha$ molecules of type $i$  with molecular volume $\nu_i$, relative to the phase volume $V^\alpha$. 
For simplicity, we consider an incompressible system where molecular volumes $\nu_i$ of each component are constants.

\begin{figure}[tb]
    \centering
      \makebox[\textwidth][c]{
      \sidesubfloat[]{\includegraphics[width=0.95\textwidth]{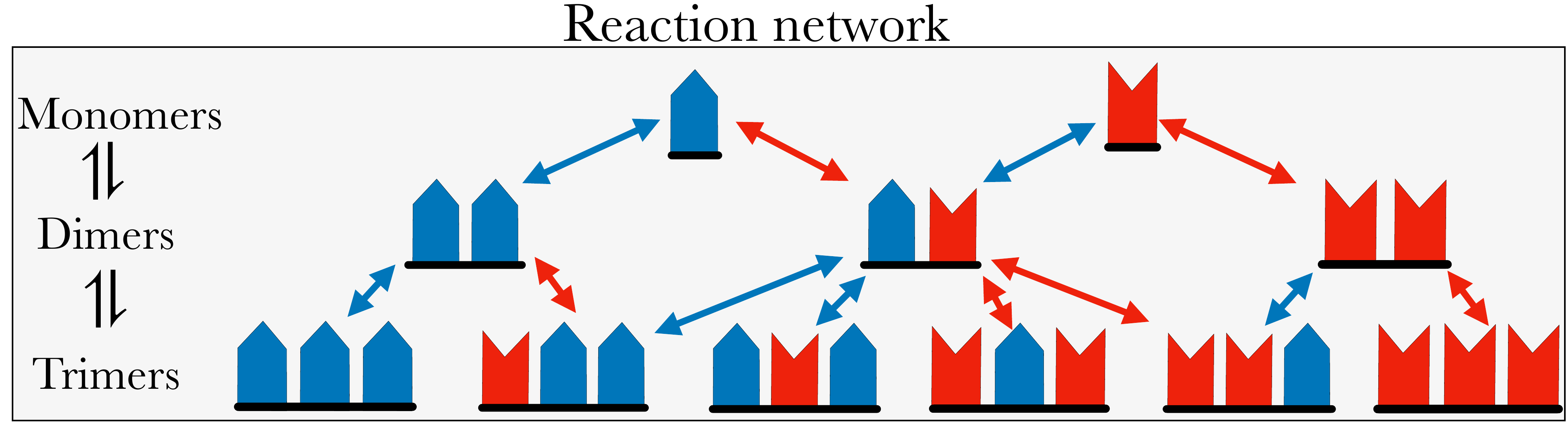}\label{fig:2a}}}\\
    \vspace{0.2cm}
      \makebox[\textwidth][c]{
      \sidesubfloat[]{\includegraphics[width=0.95\textwidth]{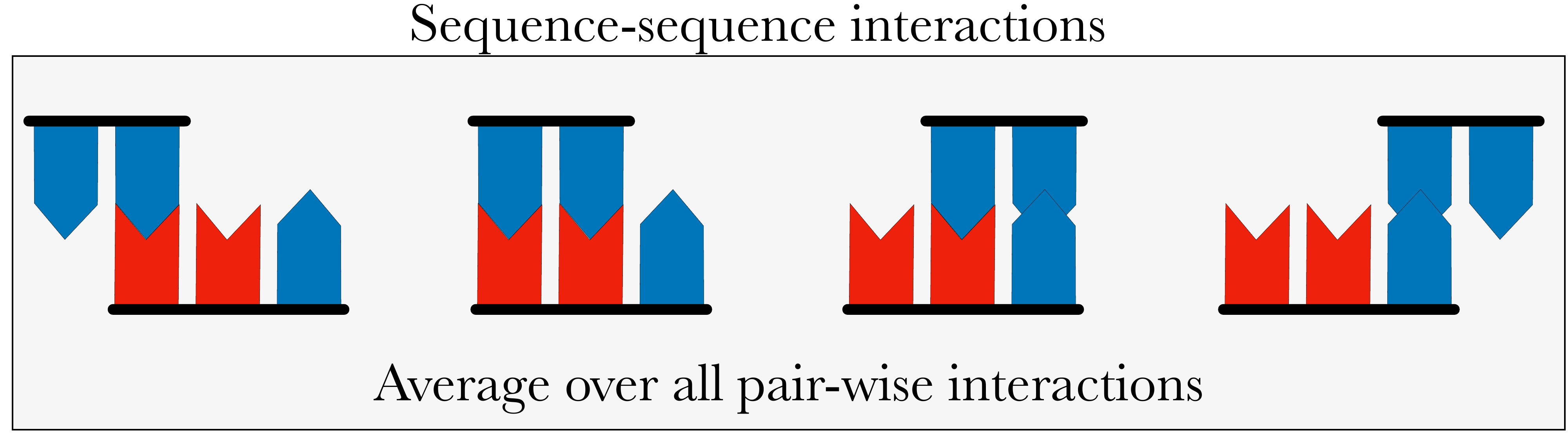}\label{fig:2b}}}\\
    \vspace{0.2cm}
      \makebox[\textwidth][c]{
      \sidesubfloat[]{\includegraphics[width=0.95\textwidth]{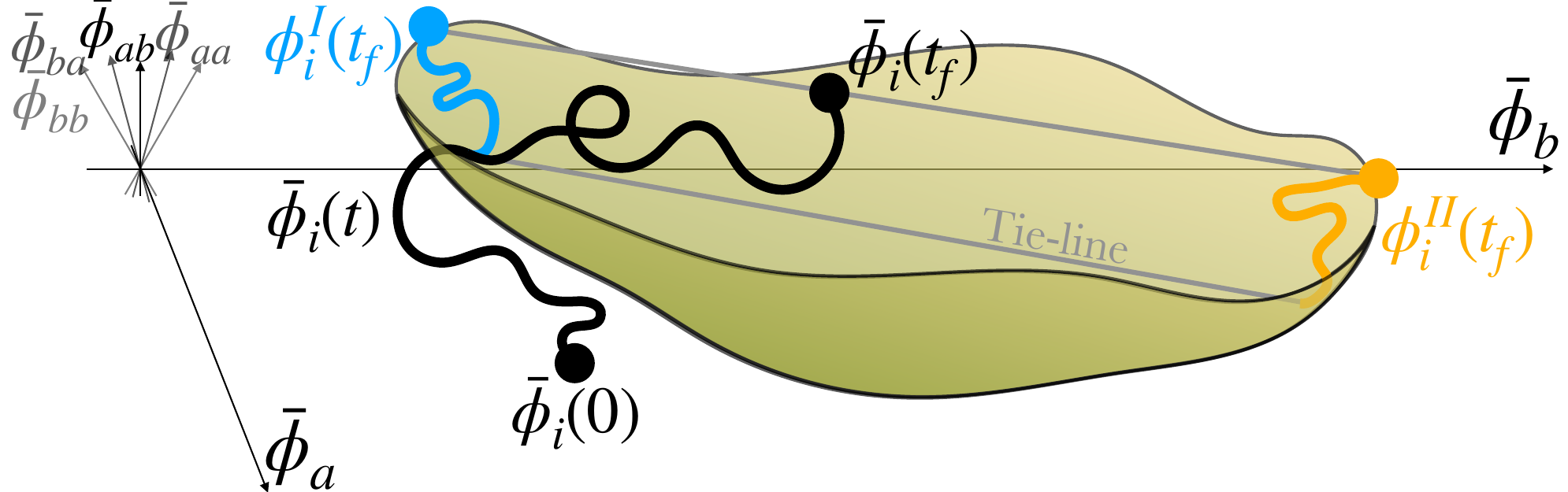}\label{fig:2c}}}
     \caption{
     \textbf{Reaction network, 
     sequence-sequence interactions, and oligomerization kinetics in the high-dimensional sequence space:} 
     (a) The reaction network for monomer pickup and release is illustrated for monomers, dimers, and trimers.
     (b) Interactions between sequences are determined by sliding over all possible pair-wise sequence configurations (Eq.~\eqref{eq:interact_case}). We average over such configurations to calculate an interaction parameter for each sequence pair  (Eq.~\eqref{eq:interact_boltzmann}). (c) Oligomerization creates a trajectory in the high-dimensional sequence space. Initializing the system in a homogeneous state with an average volume fraction $\bar{\phi}_i(t=0)$, the oligomerization trajectory  $\bar{\phi}_i(t)$ may intersect with the binodal manifold (green surface), leading to two coexisting phases. The phases I and II exhibit different compositions of oligomers (oligomer-rich and oligomer-poor) that are connected by a tie-line. 
     Phase compositions $\phi_i^\text{I/II}(t)$   move along the binodal manifold and relax to thermodynamic equilibrium or a non-equilibrium steady state (e.g., with non-equilibrium fragmentation) at time $t_{f}$.
}
\label{fig:1_react_network}
\end{figure}

In this work, we build on a theoretical framework developed for non-dilute chemical kinetics where, if phase separation occurs, the different phases $\alpha$ are at phase equilibrium~\cite{bauermann_chemical_2022}. 
In this framework, the time evolution of the components volume fractions $\phi_i^\alpha(t)$ in each phase $\alpha$ is governed by oligomerization rates $r^\alpha_{i}$, the diffusive exchange fluxes between phases $j_i^\alpha$, and changes in the phase volumes $V^\alpha$ (see Fig.~\ref{fig:1} for an illustration of the different physical processes):
\begin{subequations}\label{eq:full_model}
\begin{equation}
    \dv{t} \phi_i^\alpha = r_i^\alpha - j_i^\alpha -\frac{\phi_i^\alpha}
    {V^\alpha}
    \dv{t}  V^\alpha \, , \label{eq:kinetic_eq}
\end{equation}
with the component label $i\in\sigma$ denoting monomers and oligomer sequences. In an incompressible system, oligomerization conserves volume ($\sum_{i\in\sigma} r_i^\alpha=0$), such that the phase volume $V^\alpha(t)$ changes in time only due to partitioning fluxes of solvent $j_s^\alpha$, and partitioning fluxes of monomers and sequences $j_i^\alpha$ ($i\in \sigma$) 
through the phase boundary:
\begin{equation}
\frac{1}{V^\alpha}
\dv{t}  V^\alpha =-j_s^\alpha -\sum_{i\in\sigma}j_i^\alpha \, . \label{eq:vol_change}
\end{equation}

The oligomerization rate $r_i^\alpha$ of component $i$  results from the gain $r^\alpha_{j+m\rightleftharpoons i}$ and loss $r^\alpha_{i+m\rightleftharpoons j}$ contributions related to the possible oligomerization pathways. For simplicity, we discuss monomer pick-up and release (Fig.~\ref{fig:2a}) leading to 
\begin{equation}
\label{eq:overall_reaction_rate}
    r_i^\alpha = \sum_{j\in\sigma}\sum_{m\in\sigma_m}\left(r^\alpha_{j+m\rightleftharpoons i} - r^\alpha_{i+m\rightleftharpoons j} \right) \, .
\end{equation}
This oligomerization rate conserves the total volume (and mass) of monomers and monomers incorporated into sequences. 
In a non-dilute system, the gain and loss contribution can be written as follows~\cite{van1973nonlinear}:
\begin{equation}
    r_{i+m\rightleftharpoons j}^\alpha = k_{imj}^\alpha\left[\exp{\frac{\mu_i^\alpha + \mu_m^\alpha}{k_BT}}- \exp{\frac{\mu_j^\alpha}{k_BT}}\right] \, , \label{eq:reaction_rate}
\end{equation}
where the terms in the rectangular parenthesis denote the forward $r_{i+m\rightharpoonup j}$ and backward $r_{i+m\leftharpoondown j}$ reaction rates respectively, and $\mu_j^\alpha$ denote the chemical potential of component $j$ in phase $\alpha$. In Eq.~\eqref{eq:reaction_rate}, forward and backward rates obey
detailed balance of the rates~\cite{julicher_modeling_1997,weber_physics_2019}:
\begin{equation}
    \label{eq:det_balance}
r^\alpha_{i+m\rightharpoonup j}/r^\alpha_{i+m\leftharpoondown j} = 
\exp{\frac{\mu_i^\alpha + \mu_m^\alpha-\mu_j^\alpha}{k_BT}} \, .
\end{equation}
\end{subequations}
There is a single kinetic rate coefficient $k_{imj}^\alpha$
for each chemical reaction $i+m\rightleftharpoons j$ governing the relaxation toward chemical equilibrium, $\mu_i^\alpha + \mu_m^\alpha=\mu_j^\alpha$. 
For simplicity, we consider the kinetic rate coefficients constant, phase independent, and agnostic to the reactants, such that $k_{imj}^\alpha=k_c$. 

The non-dilute conditions require accounting for interactions among all components, the solvent, monomers, and oligomers. On a coarse-grained level, these interactions can be described by contributions to the free energy density $f(\{\phi_i \})$, which makes the chemical potential that characterizes  the free energy cost to add component $i$ to a mixture,  $
    {\mu}_{i} ( \{ \phi_j \}_{j\in\sigma}) = \nu_{i}\partial{f}/{\partial \phi_{i}}\vert_{\phi_{j\neq i}} 
  $, 
dependent on all components $j\in\sigma$. 
The free energy density $f$ contains mixing entropy contributions and enthalpic terms up to second order in each sequence volume fractions $\phi_i$ (mean-field); see, e.g., Ref.~\cite{bartolucci_interplay_2024}.
First-order enthalpic terms in $f$ correspond to internal free energies that govern oligomerization equilibria in the dilute limit.  
In our model, the internal free energies of sequences depend on sequence length, and internal free energy scales are set by backbone bond energy $\epsilon$ (Fig.~\ref{fig:1}).
We consider $\epsilon$ as sequence-independent and study values in the order of a few $k_BT$, consistent with biofilaments such as DNA and RNA~\cite{doi:10.1126/science.aaf5508}. We note that for peptides, nearest neighbor interactions along the sequences are crucial, leading to a sequence-dependent internal free energy~\cite{toal_randomizing_2015,mu_effects_2007}. 
The interactions between sequences $i$ and $j$ are captured by second-order terms in the free energy and are of the form $e_{ij} \phi_i \phi_j$, where $e_{ij}$ is the average interaction parameter. The quantity $e_{ij}$ is calculated by performing a Boltzmann average over all possible monomer-monomer interactions (corresponding to base-pairs for DNA and RNA)  between sequence $i$ and $j$, as illustrated in Fig.~\ref{fig:2b}. For details on these calculations, see Appendix~\ref{seq:F_chi_omega}.

Interactions between sequences can lead to phase separation once enough sequences have formed via oligomerization. 
If diffusion rates of sequences between phases are fast compared to oligomerization rates, 
phases are at phase equilibrium at each time of the oligomerization kinetics (see Ref.~\cite{bauermann_chemical_2022} for a detailed discussion). For two coexisting phases $\alpha=\text{I}, \text{II}$, 
chemical potentials $\mu_i^\alpha$ and osmotic pressures $\Pi^\alpha = f^\alpha - \sum_{i\in\sigma} \phi_i^\alpha\mu_i/\nu_i$ are balanced:
$\mu_i^\text{I}=\mu_i^\text{II}$ and $\Pi^\text{I}=\Pi^\text{II}$.
If not stated otherwise, we discuss scenarios where two phases coexist. 
During this kinetics, the oligomer-rich (II) and oligomer-poor (I) phases are constrained to the surface of a high-dimensional binodal manifold,  where tie-lines connect the two coexisting phases; see Fig.~\ref{fig:2c}. The constraint of phase equilibrium during oligomerization is satisfied by solving $\partial_t\mu_i^\text{I} = \partial_t\mu_i^\text{II}$ and $\partial_t \Pi^\text{I} = \partial_t\Pi^\text{II}$, and self-consistently calculating 
the diffusive fluxes between the phases, $j_i^\alpha$; for more details see Appendix~\ref{sec:appendix_BLkinetics}.
Note that the condition of phase coexistence implies the $\alpha$-superscript of the chemical potentials in Eqs. (\ref{eq:overall_reaction_rate}, \ref{eq:reaction_rate}, \ref{eq:det_balance}) can be omitted. Furthermore, as the reaction rate coefficient $k_c$ is phase independent, it implies an equal oligomerization rate $r_i$ in the oligomer-poor and -rich phase throughout the chemical kinetics.

Since chemical potentials govern both the oligomerization rates (Eq.~\eqref{eq:reaction_rate}) as well as the diffusive fluxes between the coexisting phases (last paragraph), there is a mutual coupling between oligomerization and phase separation. 
This coupling is a fundamental thermodynamic property of chemically reacting, non-dilute mixtures (also away from equilibrium); see Refs.~\cite{bauermann_chemical_2022, zwicker2022intertwined, bartolucci_interplay_2024}. In our work, it will give rise to the selective oligomerization of sequences due to their favorable interactions with their local, oligomer-rich or oligomer-poor environment.

\begin{figure*}[tb]
    \centering
     \makebox[\textwidth][c]{
     \sidesubfloat[]{\includegraphics[width=0.30\textwidth]{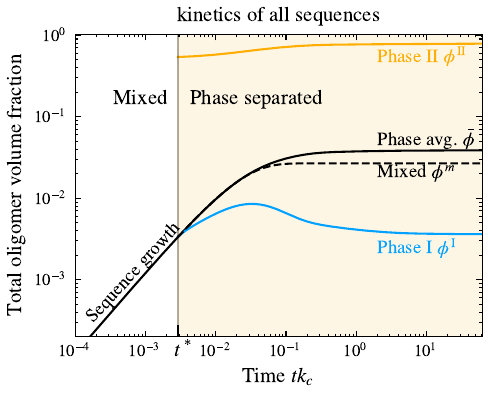}\label{fig:3a}}
     \sidesubfloat[]{\includegraphics[width=0.30\textwidth]{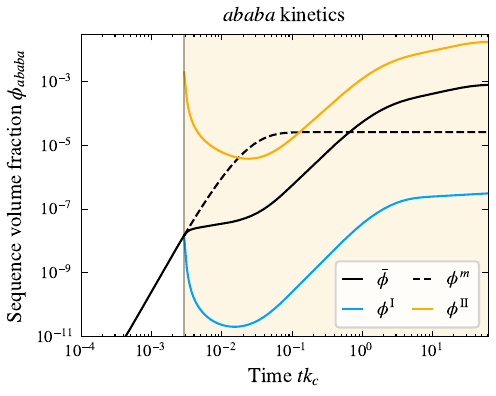}\label{fig:3b}}
     \sidesubfloat[]{\includegraphics[width=0.30\textwidth]{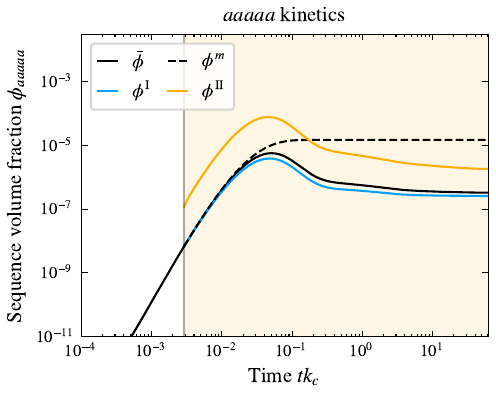}\label{fig:3c}}
      }
     \makebox[\textwidth][c]{
      \sidesubfloat[]{\includegraphics[width=1.0\textwidth]{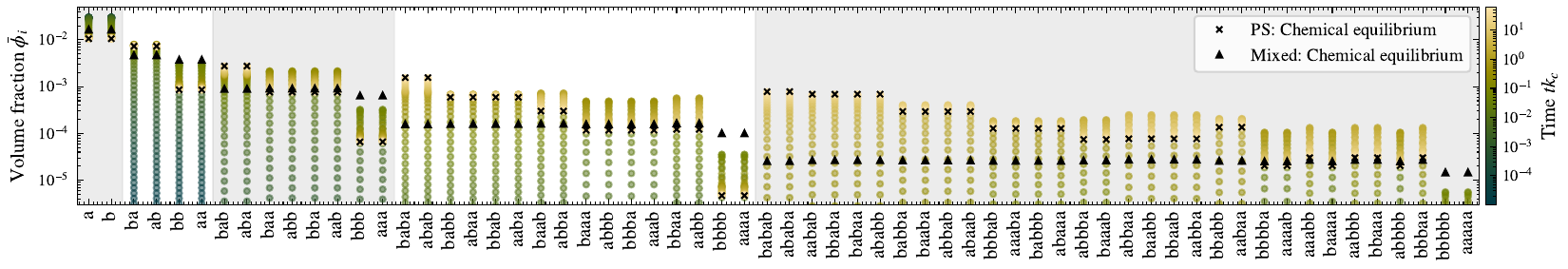}\label{fig:3d}}
      }
     \caption{
     \textbf{Oligomerization kinetics triggers phase separation leading to the enrichment of specific sequences:} 
     (a,b,c) After initializing the system with the two monomers $a$ and $b$, oligomer sequences form, increasing their volume fractions $\phi_i(t)$, causing the total oligomer volume fraction (Eq.~\eqref{eq:phi_tot_each_l}) to increase (a).
     Once sufficient oligomers have formed, the interactions among sequences trigger the formation of two coexisting, phase I (blue) and II (orange) with different volume fractions in each phase.
     After phase separation, different sequences undergo specific kinetics, leading to different phase compositions at thermodynamic equilibrium.  
     Here, we display the dynamics of the specific oligomers $i=ababa$ (b), and $i=aaaaa$ (c).
     In the phase separation domain, we also show the average volume fraction (black, Eq.~\eqref{eq:phase_avg_phi}).
     As a control, we compare it to the mixed system where phase separation is suppressed (dashed). 
     (d) The complex time evolution of the sequence distribution ranging from monomers to five-mers is generally non-monotonic, where time is indicated by the color bar. While the mixed reference case (triangle) shows an almost flat distribution for each sequence length at chemical equilibrium, there are large variations in the average volume fraction $\bar{\phi}_i(t)$ when the system can phase-separate (cross).
     }
     \label{fig:3}
\end{figure*} 

To understand the role of phase separation in the oligomerization kinetics, we consider a mixed system as a reference. The mixed system is homogeneous, and the reaction kinetics follows 
\begin{equation}
\label{eq:mixed_system}
\text{d} \phi_i/\text{d}t = r_i
\end{equation}
with the oligomerization rate given in Eqs.~\eqref{eq:overall_reaction_rate} and \eqref{eq:reaction_rate}. In simple terms, it is the same system as the phase-separated system, i.e., with the same physical parameters, but where phase coexistence is suppressed.

In the following, 
we numerically integrate Eqs.~\eqref{eq:full_model}
to evolve the volume fractions of monomers and sequences, $\phi_i^\alpha(t)$ with $i\in\sigma$ and $\alpha\in\{\RN{1},\RN{2}\}$. The average volume fractions are found from a weighted sum over both phases,
\begin{equation}
    \bar{\phi}_i = \left(V^\RN{1}\phi_i^\RN{1} + V^\RN{2}\phi_i^\RN{2} \right) V^{-1}, \label{eq:phase_avg_phi}
\end{equation}
where $V$ is the total system volume $V=V^\RN{1}+V^\RN{2}$. For all studies in this work, we initialize the system at $t=0$ with an equal amount of $a$ and $b$ monomers and no other non-solvent components. Specifically, $\bar{\phi}_a(t=0)=\bar{\phi}_b(t=0)$, and $\bar{\phi}_i(0)=0$ $\forall i\in \sigma_o$, this further sets the total sequences volume fraction $\bar{\phi}_\text{tot}=\sum_{i\in\sigma}\bar{\phi}_i$, which is conserved during the kinetics.

\section{Sequence evolution toward thermodynamic equilibrium}

\begin{figure*}[tb]
    \centering
     \makebox[\textwidth][c]{
     \includegraphics[width=\textwidth]{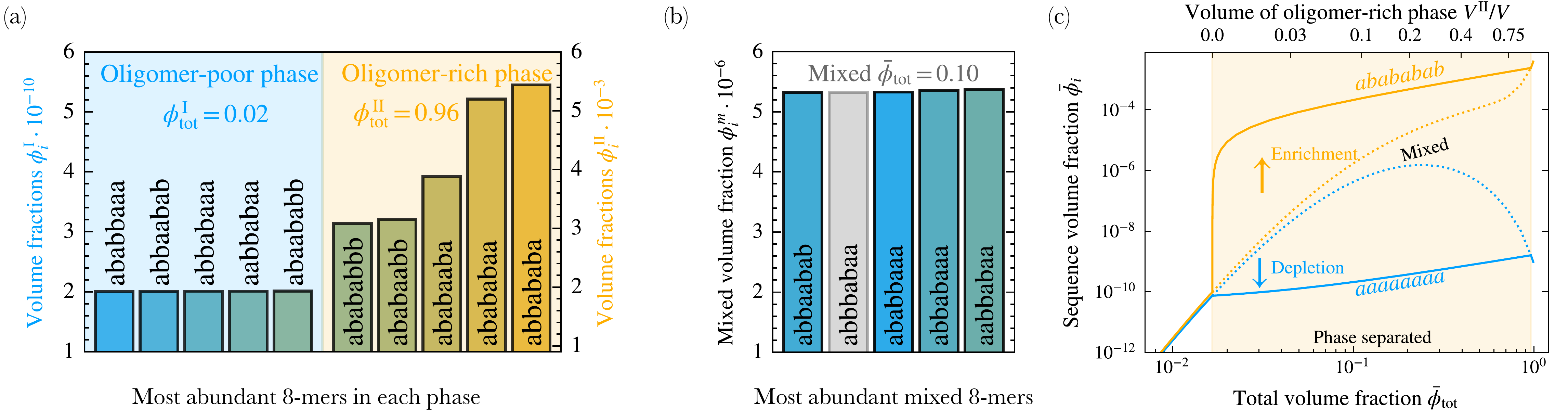}}
     \caption{\textbf{Phase separation is key for the enrichment of alternating sequences  and is most pronounced for small  oligomer-rich phases.}
     (a) The volume fraction of the five most abundant 8-mer sequences indicates that the oligomer-rich phase II biases oligomerization toward alternating sequences (within approx. one Hamming distance) by a compositional enhancement  of 
     around seven orders of magnitude relative to the oligomer-poor phase I. Moreover, in contrast to the oligomer-rich phase II,  phase I also shows no significant compositional variations among sequences. 
     (b) In the mixed system ($\bar{\phi}_\text{tot}=0.10$), the five most abundant sequences are similar to the oligomer-poor phase I and show only little compositional variations. Increasing $\bar{\phi}_\text{tot}$ toward $\phi_\text{tot}^\text{II}$, the mixed phases consistently approach the composition of the oligomer-rich phase II (see Appendix Fig.~\ref{fig:8}).
     (c) In the phase-separated regime ($\phi_\text{tot}^\text{I} <\bar{\phi}_\text{tot}<\phi_\text{tot}^\text{II}$), the relative enrichment and depletion is the largest for small volumes of the oligomer-rich phase II, $V^\text{II}$ (see upper plot axis). 
     While homopolymers (e.g., $aaaaaaaa$) get depleted at thermodynamic equilibrium, alternating sequences (e.g., $abababab$) become enriched by many orders of magnitude compared to the respective mixed cases (dotted).
     }\label{fig:4}
\end{figure*}

Dimers, trimers, and longer oligomers emerge from a pool of monomers through oligomerization. 
As more and longer oligomers form, the decrease in mixing entropy and the stronger attractive interactions enhance the propensity to phase separate. 
When phase separation occurs at $t=t^*$ (Fig.~\ref{fig:3}a), an oligomer-rich phase (phase II) forms that coexists with an oligomer-poor phase (phase I).
The two phases can differ by several orders in magnitude in their total oligomer volume fraction.
The corresponding phase average $\bar{\phi}$ typically exceeds the one of the mixed system governed by Eq.~\eqref{eq:mixed_system}, where phase separation is suppressed.

The interactions among sequences in the presence of phase separation biases the oligomerization kinetics, leading to the enrichment of specific sequences. 
As examples, we show the evolution of two different 5-mers, 
an alternating sequence $ababa$ (Fig.~\ref{fig:3}b) and
a homo-5-mer $aaaaa$ (Fig.~\ref{fig:3}c). Upon phase separation, the volume fractions in the two phases differ significantly between the two sequences. However, this difference is much more pronounced for the alternating sequence than for the homopolymeric sequence. The reason is that alternating sequences interact well with most other sequences in the system (later, we refer to such sequences as cooperative), while the homo-sequence only interacts strongly with their homopolymeric complement.
Note that it is the mutual feedback between oligomerization and phase equilibrium that biases oligomerization for more phase separation-prone sequences. 
This bias is reflected by the larger values of the average volume fractions with $\bar{\phi}_{ababa}(t)\gg \bar{\phi}_{aaaa}(t)$  for late times $t$.
It also explains why many more similarly alternating sequences form in the entire system at thermodynamic equilibrium (black crosses in Fig.~\ref{fig:3}d).

To unravel the role of the mutual feedback between phase separation and oligomerization leading to the strong enrichment of specific sequences, we considered a mixed reference system (see dashed curves in Fig.~\ref{fig:3}a-c). 
Physically, a mixed system is homogeneous, and phase separation is suppressed. Mathematically, this case corresponds to only considering the oligomerization kinetics in the non-dilute mixture by solving Eq.~\eqref{eq:mixed_system}, such that the constraint of phase equilibrium is not fulfilled.
In Fig.~\ref{fig:3}a, we see that the mixed system leads to a bit less total oligomer volume fraction $\sum_{n=2} \bar{\Phi}(n)$ \eqref{eq:phi_tot_each_l} than the phase-separated system. In other words, phase separation only slightly enhances the net trend of monomers forming oligomers. For specific sequences, however, the time evolution towards chemical equilibrium is completely altered by phase separation. This can be seen by comparing the dashed (mixed) with blue (phase I) and orange (phase II) lines in Fig.~\ref{fig:3}b-c for the two 5-mers, and when considering the full sequences distribution shown in Fig.~\ref{fig:3}d. The two sequences depicted in Fig.~\ref{fig:3b} and \ref{fig:3c} differ by less than a factor of two when mixed, while for the phase average, the same ratio is more than $10^3$. Phase separation enhances $ababa$ multiple orders of magnitude and suppresses $aaaaa$ to the extent that it is more abundant for the mixed system.
In Fig.~\ref{fig:3d}, the final chemical equilibrium volume fractions for each sequence are indicated by crosses, while  the stationary values for the mixed state are displayed by triangles. 
We observe that the mixed case has only little diversity between different sequences of the same length since it yields an almost flat sequence distribution. Note that the mixed system behaves completely different to the phase-separated case, which gives rise to specific sequences being strongly enriched while others are strongly depleted. 

A striking property of our theoretical framework is that it allows us to follow the  kinetics of the full sequence distribution in time (Fig.~\ref{fig:3}d).
Longer oligomers form at the expense of depleting the amount of initialized monomers.  At early times, the kinetics is dominated  by the entropic gain to distribute the monomer mass to larger oligomers.
This fast entropy-driven  growth of oligomer can lead to overshoots in the volume fractions at intermediate times when sequences are abundant enough such that interactions begin to affect the oligomerization kinetics  (see e.g., Fig.~\ref{fig:3}c). 
At  later times,  the  enthalpic gains through interactions balance with the entropic costs of unequally distributed mass between sequences of the same length. 
In the mixed case, the system is not dense enough for the interactions to alter the chemical kinetics significantly, resulting in a simple monotonic increase in the volume fractions (dashed lines in Fig.~\ref{fig:3}b and c).  Eventually, thermodynamic equilibrium is reached where phase coexistence and chemical equilibrium are concomitantly satisfied, and the sequence distribution in each phase becomes stationary (black crosses in  Fig.~\ref{fig:3}d). At thermodynamic equilibrium, the  most and least abundant sequences among 5-mers differ in volume fraction by more than three orders of magnitude, as shown in Fig.~\ref{fig:3}b and c. Phase separation thus strongly alters kinetics and the sequence distribution at thermodynamic  equilibrium.

\begin{figure*}[tb]
    \centering
    \makebox[\textwidth][c]{
   {\includegraphics[width=\textwidth]{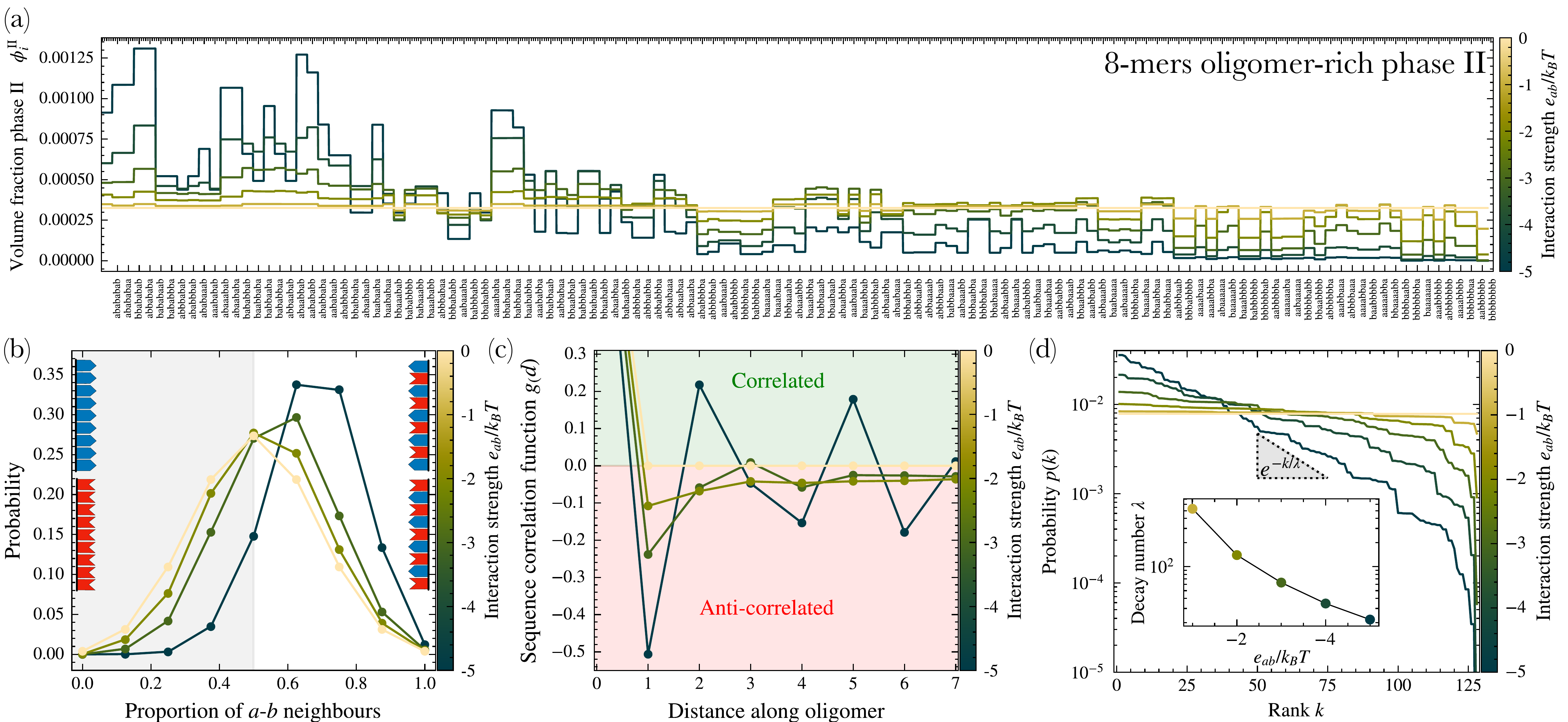}}}
     \caption{
     \textbf{
Formation of alternating sequences requires phase separation and 
     strong enough sequence interactions. }
     (a) There are strong variations between the equilibrium volume fraction of sequence in the distributions of 8-mers in the oligomer-rich phase II. The enrichment or depletion of specific sequences originates from sequence-sequence interactions $e_{ab}$ (values indicated by color code).
    (b,c) Alternating sequences require strong enough attractive interactions, $e_{ab}$ 
    This trend is reflected in (b), showing the $a$-$b$ neighbor distribution getting more skewed toward more $a$-$b$ neighbors, and (c) with the sequence correlation function $g(d)$ (Eq.~\eqref{eq:correlation_length}) developing a clear oscillatory behavior with a wave number of about two monomers (i.e., units of $ab$). 
    (d) Ordering the sequences by abundance gives the rank distribution $p(k)$, which decays exponentially,  $p\propto e^{-k/\lambda}$. The characteristic rank scale $\lambda$ decreases with more adhesive interactions $e_{ab}$. This behavior indicates that more monomers are shuffled to a smaller amount of sequences to enhance their abundance. In other words, variations in the sequence distribution enhance.
    }
\label{fig:5}
\end{figure*}

The key property of phase separation is that the coexisting phases act as local hubs with very different total oligomer mass and, thereby, very different inter-sequence interactions (Fig.~\ref{fig:3}a). 
These interactions affect oligomerization  locally, enriching or depleting specific sequences.
This hub-like effect is evident by comparing the five most abundant sequences in each phase (Fig.~\ref{fig:4}a). While the oligomer-poor (I) phase has no clear preference for any sequence (see also Fig.~\ref{fig:5_appendix}a), the oligomer-rich (II) phase shows a clear preference towards alternating sequences, each of varying abundance. For the corresponding mixed system with a total volume fraction $\bar{\phi}_\text{tot}=0.10$ (Fig.~\ref{fig:4}b), the distribution is almost flat, similar to the oligomer-poor phase. Most importantly, phase separation has a dual effect: It changes which sequences are enriched and depleted and also magnifies their depletion and enrichment. This can be seen by sequences in the oligomer-rich (II) phase (indicated by orangish color code) that do not appear in either the oligomer-poor (I) or the mixed case. 
Consistently, the oligomer-poor (I) phase and the mixed system at low total oligomer volume fraction enrich similar sequences  such as $aabbabaa$.

The enrichment mediated by phase separation is most pronounced for small  volumes of the oligomer-rich phase II. This property is apparent by comparing the equilibrium volume fraction averaged over both phases (Eq.~\eqref{eq:phase_avg_phi}) with the mixed case where phase separation is suppressed, for varying the total volume fraction of monomers and  oligomers $\bar{\phi}_\text{tot}$; see Fig.~\ref{fig:4}c. Here, phase separation abruptly enriches the volume fraction of the alternating 8-mers, while depleting the homo-8-mers. The relative enrichment is the largest for small volumes of the oligomer-rich phase. 
This behavior is due to the volume fractions of long sequences increasing exponentially with the total mass (Eq.~\eqref{eq:equil_phi}), while the phase average increases linearly with the composition of the oligomer-rich phase (Eq.~\eqref{eq:phase_avg_phi}).
As a result, there is a relative enrichment of specific sequences up to a thousand-fold. For the mixed system, the two sequences depicted remain at approximately equal abundance until around $\phi_\text{tot}\simeq 0.1$, while phase separation is able to differentiate them at more dilute conditions. Phase separation thus allows for the enrichment of specific sequences at  conditions slightly above the saturation volume fraction, which would not be possible without phase separation. 

Sequence interactions collectively enrich similar sequence patterns in the respective phases. 
In the oligomer-poor phase, the oligomers are too dilute to affect each other's propensity through interactions, as seen in Fig.~\ref{fig:5_appendix}a and Eq.~\eqref{eq:rel_prop}.
In the oligomer-rich phase, the oligomer-oligomer interactions determine the sequence distribution, as seen in Fig.~\ref{fig:5_appendix}b. Without interactions ($e_{ab}=0$) or a too-weak interaction strength $e_{ab}$, the distribution remains approximately flat for each length, independent of the total volume fraction of monomers and sequences $\phi_\text{tot}$. With stronger interactions $e_{ab}$, the volume fractions of the enriched/depleted sequences may enhance/deplete further (see Fig.~\ref{fig:5}a for 8-mers).
However, due to the dual role of phase separation creating a sequence bias and acting as an amplifier, a non-monotonous behavior can be observed for some sequences when increasing the interaction strength $e_{ab}$.  For example,  $abaababbb$ is enriched at $e_{ab}=-4k_BT$, but depleted at $-5k_BT$. Another example is $aabababaa$ being the most abundant sequence at $e_{ab}=-3k_BT$, but that is far from the case at $-5k_BT$. These non-monotonous trends result from enrichment being a collective effect, where the enhanced depletion of $aa$ and $bb$ neighbors by interactions has an effect on the abundance of all sequences. 

Sequence interactions and phase separation mediate a strong bias toward sequences with almost alternating sequence patterns at thermodynamic equilibrium (Fig.~\ref{fig:4}a Fig.~\ref{fig:5}a).
This behavior is evident in the probability distribution for the proportion of $a$-$b$ neighbors for 8-mers, where $0$ corresponds to a homo-oligomer and $1$ corresponds to an alternating oligomer (Fig.~\ref{fig:5}b). Without interactions, the distribution is a binomial distribution symmetric around an $a$-$b$ neighbor proportion of a half. With stronger interactions $e_{ab}$, the mean of the distribution shifts to a higher number of $a$-$b$ neighbors, and the variance of the distribution decreases. The results support that oligomers close to homo-oligomers are statistically disfavoured as they have fewer complementary strand motifs where sequences can attach. Due to the volume fractions decaying with length, the interactions with shorter sequences are the most relevant for enhancing motifs in longer sequences. Though alternating patterns are preferred, a perfect alternating pattern does not produce any complementary sub-strand for other non-perfectly alternating sequences and is, hence, not as enriched. The largest enrichment occurs for approximately alternating sequences that interact favorably with $aa$ or $bb$ segments of other sequences. 

Phase separation and interactions among oligomers create long-range correlations along the oligomer sequence. The corresponding auto-correlation function 
\begin{equation}
    g(d) = \frac{\sum_{i\in\sigma}\delta_{n_i,l}\sum_{j=1}^l \phi_i\sum_{\pm}\begin{cases}
    1 \,\,\,\,\,\,\,\text{if $\,\,\Gamma_{i,j\pm d}=\Gamma_{i,j}$} \\
    -1 \,\,\,\text{if $\,\,\Gamma_{i,j\pm d}\neq\Gamma_{i,j}$}
    \end{cases}}{2l\sum_{i\in\sigma}\phi_i\delta_{n_i,l}} , \label{eq:correlation_length}
\end{equation}
characterizes the occurrence of the same monomers along an oligomer of fixed length $l$. In the equation above,  $\Gamma_{i,j}$ gives the monomer-type at position $j$ of oligomer $i$.
The auto-correlation function $g(d)$ shows pronounced oscillations  for large enough interaction strength $e_{ab}$ (Fig.~\ref{fig:5}c). Stronger interactions (more negative values of $e_{ab}$) give rise to stronger correlations, while no interactions ($e_{ab}=0$) leave all sequences uncorrelated.
The wave number of the correlation function $g(d)$ is approximately two, corresponding to the alternating pattern $ab$, which is statistically most common.

The stronger sequences interact (more negative $e_{ab}$), the more of the total monomer mass gets turned into enriched sequences. This trend can be quantified by the rank distribution $p(k)$ that 
orders the sequences by their abundance (rank $k$). We find that the rank distribution $p(k)\propto \exp(-k/\lambda)$ decays exponentially,
with the characteristic rank scale $\lambda$ decreases with the strength of interaction $e_{ab}$. Without interactions ($e_{ab}=0$), the sequence distribution for a given length is flat, and so is the rank distribution $p(k)$ (yellow line shown in Fig.~\ref{fig:5}d). 
With increasing interaction strength, some  sequences get  enriched while most are depleted. The characteristic rank scale $\lambda$ thus decreases 
with interaction strength $e_{ab}$ (inset of Fig.~\ref{fig:5}d). As most sequences become depleted at large interaction strengths $e_{ab}$, interactions  reduce the size of the effectively occupied high-dimension sequence space.

The preference towards alternating sequences originates from the necessity of all sequences to interact with both $a$'s and $b$'s. 
The preference for alternating sequences
results from the interaction strength $e_{ab}$ 
and the volume fractions of all sequences being large enough, which both  affect the distribution 
in an  exponential fashion
(Eq.~\eqref{eq:equil_phi_es}).
While large sequences are dilute compared to short oligomers (dimers, trimers, ...), the dominant contributions in the exponential leading to the enrichment of alternating patterns in long oligomers (7-, 8-, 9-mers), stem from short oligomers (see inset Fig.~\ref{fig:5_appendix}b). 
The origin of short alternating sequences arises from them interacting more favorably with their environment, meaning their respective phases, i.e., they are more cooperative.  
For example, the dimer $ab$ is much more abundant than $aa$ (and $bb$) as it interacts more favorably with 
both monomers $a$ and $b$, see Fig.~\ref{fig:3d} and \ref{fig:5_appendix}b. 
But why are the most enriched sequences approximately one Hamming distance away from the perfectly alternating 8-mer $ababababab$, and not the perfectly alternating 8-mer itself? 
The reason is that a single non-alternating segment,
$aa$ or $bb$, enhances the average interactions with the diverse sequence environment, further boosting a sequence's cooperativity.

\section{Sequence selection by non-equilibrium fragmentation}

\begin{figure}[tb]
    \centering 
    \makebox[\textwidth][c]{        \includegraphics[width=\textwidth]{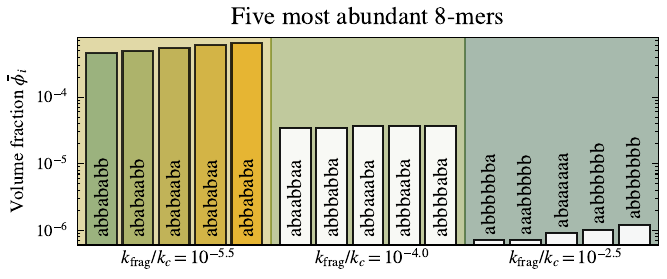}}
     \caption{
     \textbf{Five most abundant sequences when phase-separation is maintained away from equilibrium by fragmentation:}
    With increasing fragmentation rate coefficient $k_\text{frag}$, the selection bias changes from alternating sequences to sequences with longer blocks of the same monomer type. For very large  $k_\text{frag}$, these blocks approach the maximal sequence length, and the most abundant sequences deviate from homopolymers within 1-2 Hamming distances.}
     \label{fig:5.5}
\end{figure}

\begin{figure*}[tb]
    \centering 
    \makebox[\textwidth][c]{        \includegraphics[width=\textwidth]{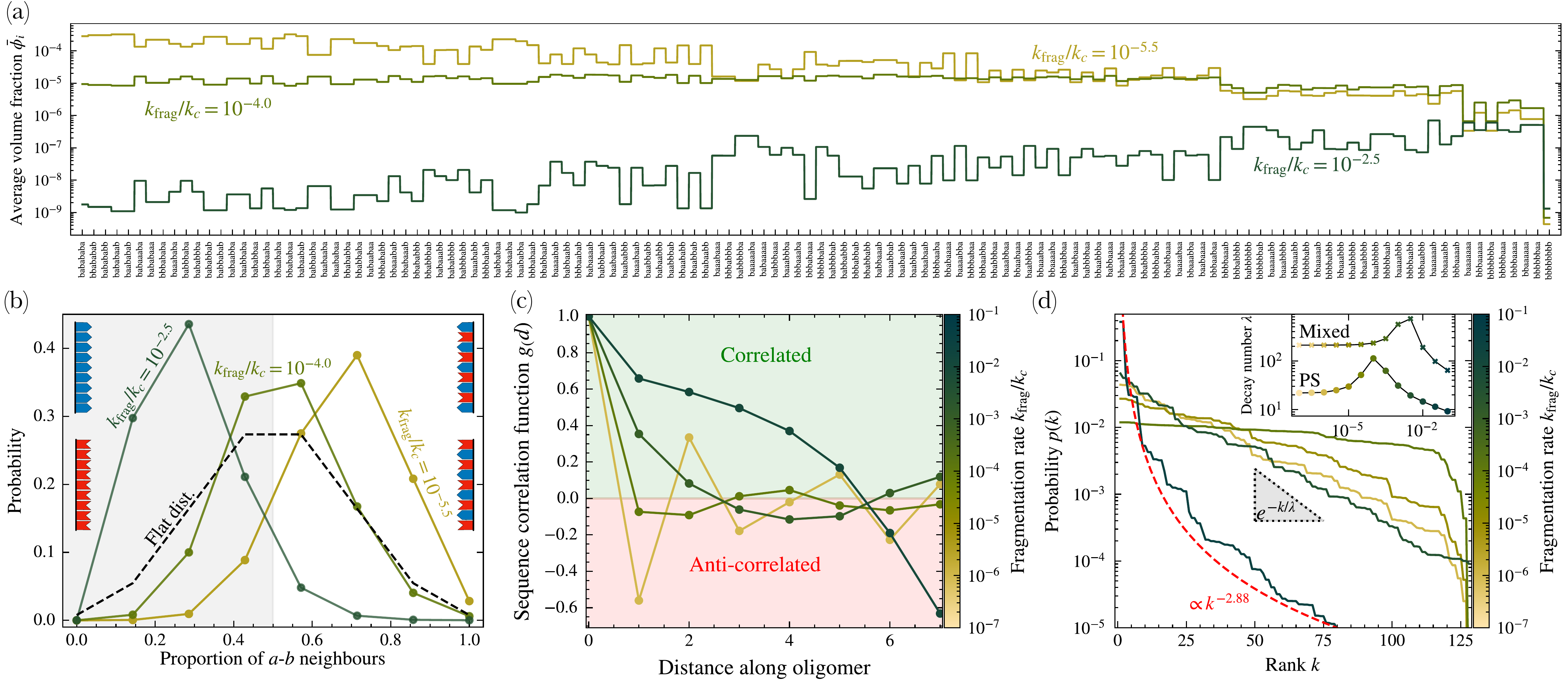}}
     \caption{
     \textbf{Non-equilibrium fragmentation creates a selection pressure favoring sequences with longer blocks of the same monomer type:}
    (a,b) With increasing fragmentation rate coefficient $k_\text{frag}$, the selection bias changes from alternating sequences to sequences with longer blocks of the same monomer type. For very large  $k_\text{frag}$, 
    these blocks approach the maximal sequence length (8mers in (a)), and the most abundant sequences deviate from homopolymers within 1-2 Hamming distances.
    (b,c) The selection trend for increasing $k_\text{frag}$ from alternating sequences to sequences with larger blocks is supported by (b) showing a pronounced shift toward less $a$-$b$-neighbors and a pronounced increase of the wave number of the sequence correlation function correlation function $g(d)$ (see (c)). 
    (d) The rank distribution $p(k)\propto \exp(-k/\lambda)$ decays exponentially  for smaller fragmentation rate coefficients $k_\text{frag}$. 
    The  characteristic rank scale $\lambda$ (inset) has a non-monotonous behavior and is maximal for intermediate values of $k_\text{frag}$ because the sequence distribution is approximately flat (a).  For the very large fragmentation rate coefficients $k_\text{frag}$, the rank distribution follows a power-law decay with $p(k)\propto k^{-\beta}$ with an exponent $\beta\simeq 2.88$.}
     \label{fig:6}
\end{figure*}

To decipher the effects of phase separation in sequence selection away from equilibrium, we break detailed balance of the rates between the fragmentation and fusion pathways $i\rightharpoonup j + \abs{i-j}$. 
In general, both pathways rely on the break-up (formation) of bonds along the sequence backbone, where sequence $i$ fragments into (forms from) 
the sub-oligomer/monomer
$j$  and $\abs{i-j}$.
An example is $aabab\rightharpoonup aa +bab$.
Detailed balance of the rates is broken by omitting the fusion pathway, considering it to be slow compared to fragmentation. Thus, the system is maintained away from equilibrium. The rate of the fragmentation pathway for sequence $i$ in phase $\alpha$ is written as
\begin{equation}
    h_{i\rightharpoonup j+\abs{i-j}}^\alpha = k_\text{frag}\frac{\phi_s^\alpha\phi_i^\alpha}{n_i-1} \, , \label{eq:hydro_cut}
\end{equation}
where $k_\text{frag}$ denotes the fragmentation rate coefficient that we choose, for simplicity, to be phase-independent.
The  fragmentation rate $h_{i\rightharpoonup j+\abs{i-j}}^\alpha$ is proportional to the local  volume fraction $\phi_i^\alpha$ ($\alpha=\text{I,II}$) of the fragmenting sequence. Each bond along the sequence backbone breaks equally likely, making the rate independent of $j$, ensured by the factor $(n_i-1)^{-1}$, where $n_i$ is the number of monomers of sequence $i$. 
Inspired by the hydrolysis of biofilaments such as DNA and RNA, the fragmentation rate scales with the solvent volume fraction $\phi_s^\alpha$. 

Breaking detailed balance of the rates in the oligomerization kinetics gives rise to a  selection pressure that strongly changes the sequence distribution compared to  thermodynamic equilibrium. 
For the parameters studied, we always observe a non-equilibrium steady state with a stationary sequence distribution. For low fragmentation rate coefficients $k_\text{frag}$,  alternating sequences are favored, and the sequence distribution remains almost unchanged compared to thermodynamic equilibrium, as shown in Fig.~\ref{fig:5.5} and the yellow distribution in Fig.~\ref{fig:6}a. 
For intermediate values values of $k_\text{frag}$ (light green), the thermodynamic sequence bias is no longer evident, and the non-equilibrium sequence distribution at steady state is almost flat with approximately equal volume fractions for each sequence.  For very large fragmentation rate coefficients $k_\text{frag}$,
we find that the thermodynamic bias of sequences observed for vanishing $k_\text{frag}$  inverts, i.e.,  
sequences that are most depleted at thermodynamic equilibrium become most abundant in the non-equilibrium steady state, and vice versa. 

The inverted thermodynamic bias for large fragmentation rate coefficients $k_\text{frag}$ originates from thermodynamically unfavorable oligomers having the largest oligomerization rates. 
%
The inverted trend is related to the backward oligomerization pathway of monomer release being negligible compared to the fragmentation pathway for large $k_\text{frag}$. 
The result is, following the derivation in Appendix \ref{sec:app_NEQ}, a non-equilibrium steady state (NESS) sequence distribution in the oligomer-rich phase (II): 
\begin{equation}
    \phi_i^{\RN{2}, \text{NESS}} \simeq \frac{k_c}{k_\text{frag}}\frac{\exp\left(\frac{\mu_m}{k_BT}\right)}{\frac{V^{\RN{2}}}{V}\phi_s^{\RN{2}}}\sum_{j\in\mathcal{R}_i}\exp\left(\frac{\mu_j}{k_BT}\right) \, ,\label{eq:approx_NESS_dist}
\end{equation}
where $\mu_m$ is the monomer chemical potential (Eq.~\eqref{eq:chem_pot}).
We note that the effects of other sequences $j$ on the steady state abundance of sequence $i$ is described by the term $\sum_{j\in\mathcal{R}_i}\exp\left({\mu_j}/{(k_BT)}\right)$.  
Thus, a sequence $i$ is more abundant if the set of sequences that oligomerize into $i$ ($j\in\mathcal{R}_i$) have unfavorable interactions with their environment, reflected in a larger value of their chemical potential $\mu_j$.
In other words, non-equilibrium fragmentation selects for sequences of lower cooperativity, that are homo-oligomer-like sequences. 
This selection mechanism is the opposite of the thermodynamic enrichment mechanism that favors sequences of higher cooperativity.

For increasing fragmentation rate coefficient $k_\text{frag}$, the most abundant sequences change from alternating oligomers to oligomers containing low complexity domains composed of the same monomeric unit.
This trend is illustrated by the five most abundant sequences (Fig.~\ref{fig:5.5}) and by the probability distribution of $a$-$b$-neighbors (Fig.~\ref{fig:6}b). We see that the five most abundant sequences for weak non-equilibrium driving ($k_\text{frag}/k_c=10^{-5.5}$) are alternating sequences with maximally two monomers of the same type neighboring each other. In this case, the $a$-$b$-neighbor distribution is skewed toward an $a$-$b$-proportion of 1, indicating that most $a$-monomers are neighbored by $b$ monomers. Increasing $ k_\text{frag}$ to intermediate values enhances the length of domains composed of the same monomer type. For example, the five most abundant sequences contain sequences similar to $abbbaba$, containing both alternating and homo-oligomeric parts. This results in an $n_{a,b}$-distribution peaked around a half, but less spread than the flat distribution. 
When increasing the fragmentation rate coefficient $k_\text{frag}$ further, 
domains composed of the same monomer type approach the max length, leading to almost homopolymeric sequences such as $abbbbbbb$ or $abaaaaaa$. Consistently, the $a$-$b$-distribution is more skewed toward a small fraction of $a$-$b$-neighbors for increasing $k_\text{frag}$.   
To quantify the shift from alternating to sequences with more extended 
domains of the same monomer type, we use the correlation function defined in Eq.~\eqref{eq:correlation_length}.
Fig.~\ref{fig:6}c confirms that faster fragmentation leads to an increase in the wave number of the correlation function, corresponding to longer correlated domains of the same monomer type.

For weak non-equilibrium driving the exponential decay of the rank distribution of 8-mers, $p(k)\simeq \exp{(-k/\lambda)}$ persists, as displayed in Fig.~\ref{fig:6}d. However, for intermediate fragmentation rates $k_\text{frag}$, the characteristic decay length $\lambda$ increases to a maximum  before decreasing. At the maximum, the sequence distribution is almost flat. 
A steeper rank distribution (smaller $\lambda$) indicates a stronger selection pressure, where only a small subset of the sequences are selected. 
For very large fragmentation rate coefficients $k_\text{frag}$, the rank distribution approximately follows a power-law $p(k)\propto k^{-\beta}$ with $\beta\simeq 3$. 
Power-laws in rank distributions were reported for many complex systems such as city sizes, gene circuits, and self-replicating systems~\cite{sole_zipflaw2002}.  
Though the exponent is different from Zipf's law~\cite{Zipf+1932}, our finding suggests that our phase-separated-system with non-equilibrium oligomerization is complex enough for enabling open-ended evolution~\cite{sole_zipflaw2002}.
We note that without phase separation (mixed case), there is also a power-law behavior of the rank distribution for very large $ k_\text{frag}$ with an exponent close to Zipf's law (see Appendix, Fig.~\ref{fig:10}).

\begin{figure*}[tb]
    \centering
     \makebox[\textwidth][c]{
    \includegraphics[width=\textwidth]{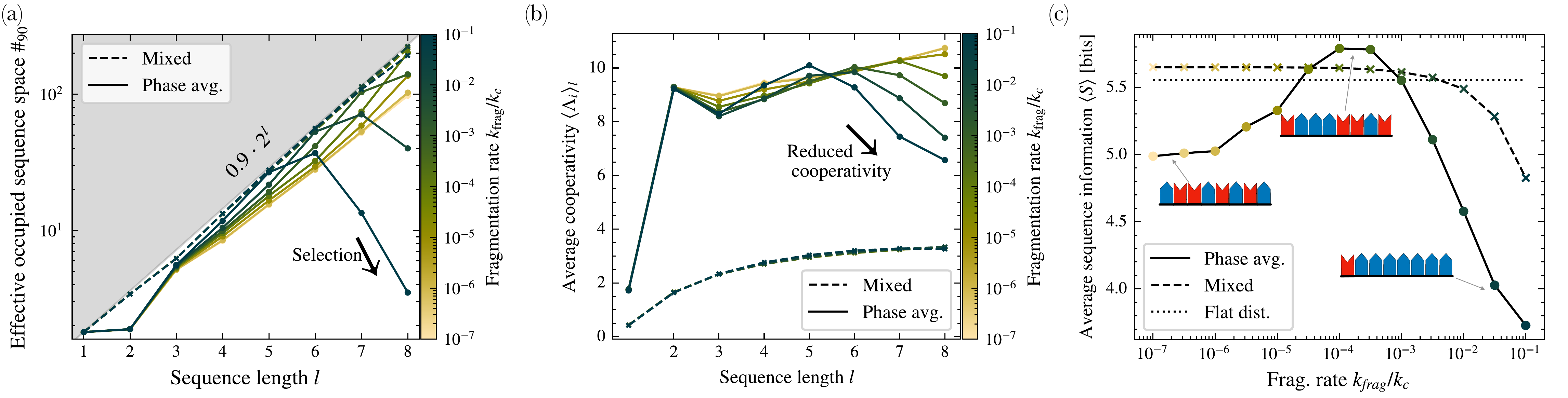}}
     \caption{
     \textbf{Non-equilibrium fragmentation selects for less cooperative sequences of lower information.} 
     (a) The number of sequences $\#_{90}(l)$ occupying $90\%$ of the oligomer volume of a given length $l$ (Eq.~\eqref{eq:etta_def}) deviates from the exponential scaling as observed at thermodynamic equilibrium (vanishing values of  $k_\text{frag}$). 
    For the large fragmentation rate coefficients $k_\text{frag}$, $\#_{90}(l)$ even decreases for long enough sequences, indicating a strong selection pressure selecting  a sub-exponential subset of the possible sequence for a given length $l$.
    (b) The cooperativity characterizes how well sequences of each length interact with their environment (Eq.~\eqref{eq:fitness}). Thermodynamics promotes cooperativity (low $k_\text{frag}$) and increases with sequence length $l$ as the sequences have more freedom to adapt their structure to suit their environment. Increasing the fragmentation rate coefficient $ k_\text{frag}$ changes the selection pressure to promote oligomers with uncooperative substrates, reducing the average cooperativity at large $l$.
    (c) The average sequence information over the distribution of 8-mers \eqref{eq:SMI} has a local maximum with the fragmentation rate coefficient $k_\text{frag}$. In the limit of low and high fragmentation rates, the average information content of the sequence distribution decreases due to the selection of alternating sequences or sequences of larger homopolymeric blocks, respectively.
     Similar behavior is found for different measures of sequence information/complexity, such as the  Kolmogorov complexity~\cite{sole_zipflaw2002} and the Lempel–Ziv complexity~\cite{lempel_ziv}.
     }
     \label{fig:7}
\end{figure*}

Maintaining oligomerization away from equilibrium gives rise to a strong selection mechanism, where only a small subset of all possible sequences accumulate most of the total oligomer volume fraction for that length. This effect is most pronounced for longer sequences. 
We characterize these observations by the quantity $\#_{90}(l)$  that is the smallest number of sequences corresponding to $90 \%$ volume fraction for a given length $l$ (Fig.~\ref{fig:7}a): 
\begin{equation}   \frac{\sum_{i\in\sigma}^{\#_{90}(l)}\phi_i\delta_{n_i,l}}{\sum_{i\in\sigma}\phi_i\delta_{n_i,l}} > 0.90 \, . \label{eq:etta_def}
\end{equation}
This quantity characterizes the effectively occupied sequence space. 
At thermodynamic equilibrium, and independent of the interaction strength $e_{ab}$, $\#_{90}$ scales exponentially in sequence length because longer length leads to exponentially more sequences (see Fig.~\ref{fig:7}a for low $k_\text{frag}$). Strikingly, the selection pressure, once driven away from equilibrium, gets so strong for fast enough fragmentation (large $k_\text{frag}$) that 
$\#_{90}(l)$ scales even sub-exponentially (Fig.~\ref{fig:7}a). Though the sequence distribution is less diverse when maintaining oligomerization away from equilibrium, the selection pressure facilitates specific sequence motifs to win over the others.
The findings of sub-exponential scaling exemplify the potential of phase separation with oligomerization maintained away from equilibrium in providing a selection mechanism for short oligomers as they explore the exponentially large space of possible sequences.

Selection mediated by phase separation can be further corroborated  by introducing a cooperativity measure. The cooperativity per sequence length ($n_i$) $\Lambda_i$ measures how well sequence $i$ interacts with its environment: 
\begin{equation}
    \Lambda_i = \gamma_i^{-n_i}\left( \{\phi\}\right)\exp{-\frac{\mu_i^0}{n_i\,k_BT}}\, . \label{eq:fitness}
\end{equation}
Here, the chemical potentials have been decomposed in terms of the composition-independent reference chemical potential $\mu_i^0$, and the composition-dependent activity coefficient $\gamma_i$, such that $\mu_i = \mu_i^0 + k_BT\log{\left(\phi_i\, \gamma_i\right)}$, with the expressions given in Eq.~\eqref{eq:chem_pot_decomp}. Chemical equilibrium enriches oligomers that interact well with sequences in their respective phase, corresponding to a larger  average cooperativity  (Fig.~\ref{fig:7}b). The longer an oligomer, the more cooperative segments exist at which other sequences can attach. 
We note that without phase separation (mixed case), sequences are relatively uncooperative and not well-adapted to their mixed environment. 
The increasing cooperativity trend with sequence length in the phase-separated case at thermodynamic equilibrium changes when maintaining oligomerization away from equilibrium. 
The cooperativity decreases as the selection pressure inverts the thermodynamic enrichment, leading to less alternating and more homopolymeric sequences. In other words, the faster the fragmentation, the more uncooperative sequences are selected. A selection for less cooperative sequences is a prerequisite for a prebiotic system to develop specific functionalities that are otherwise suppressed at thermodynamic equilibrium.

Finally, we ask whether the selected sequences store more or less information. 
We quantify the information content of each sequence $S_i$ 
by the number of $a$-$b$-neighbors
using  Shannon's measure of information~\cite{shanon_info, ben2017modern}:
\begin{equation}
\label{eq:SMI}
     S_i= -\sum_{j\in\sigma} p(j|n_{a,b}^{(i)}) \log_2{ \left( p(j|n_{a,b}^{(i)}) \right)} \, ,
\end{equation}
where $p(j|n_{a,b}^{(i)})$ is the probability of sequence $j$ given the number of $a$-$b$-neighbors $n_{a,b}^{(i)}$ of sequence $i$, without prior knowledge about the sequence distribution. No prior knowledge corresponds to a flat sequence distribution.
This means that each sequence of the given length with a specific $n_{a,b}^{(i)}$ value is equally likely. In this case, Eq.~\eqref{eq:SMI}
gives the maximal 
value of
\begin{equation}
S_{i}^*=-\log_2{ \{ p(i|n_{a,b}^{(i)}))\}} \, .
\end{equation} 
Thus, $S_{i}^*$ quantifies the information gained in bits by knowing the exact sequence, with the only prior information being the number of $a$-$b$ neighbors corresponding to sequence $i$.
For example, there are only two sequences of length $l=8$ that are homo-oligomers $(n_{a,b}=0)$. Thus, the probability of picking the correct homo-oligomer is one-half, yielding an information gain of 1 bit by finding out the actual sequence. Similarly, the sequence $abbbaaba$ has a probability of $1.4\%$ among all sequences of four $a$-$b$-neighbors, yielding an information gain of $6.1$ bits. Thus, $S_{i}^*$  assigns the same value of information to the sequences with the same value of $n_{a,b}$. 
The average information of a sequence distribution is calculated as $\langle S \rangle = \sum_{i\in\sigma}\delta_{n_i,l}\bar{\phi}_iS_i^*/\sum_{i\in\sigma}\delta_{n_i,l}\bar{\phi}_i$.

Phase-separated systems away from equilibrium change the information content of the selected sequences. 
Fig.~\ref{fig:7}c shows the average sequence information of 8-mers, $\langle S \rangle$, at the non-equilibrium steady state for increasing non-equilibrium fragmentation rate coefficient $k_\text{frag}$. For phase-separated systems, $\langle S \rangle$ has a non-monotonous dependency on $k_\text{frag}$. Close to thermodynamic equilibrium (low $k_\text{frag}$), phase separation favors low-information sequences, i.e., alternating sequences. Increasing the fragmentation rate coefficient $k_\text{frag}$ leads to an increase in the average sequence information, exceeding the corresponding average for a flat distribution of 8-mers. The reason is that there is an enrichment of sequences with $3$ and $4$ $a$-$b$-neighbors at the expense of homo-oligomers and alternating oligomers (Fig.~\ref{fig:6}b), which contain less information. Note that no such local maxima is observed without phase separation, where a monotonic decrease in average information content with fragmentation rate $k_\text{frag}$ is observed. Increasing $k_\text{frag}$ further, in the case with phase separation, selects for longer blocks of the same monomer type, strongly decreasing the average information content. 
Such low-information sequences are particularly interesting because they have the potential to show rich functionalities, such as catalytic activity or folding, and the extended blocks can mediate specific, lock-like interactions.

\section{Conclusion}


In our work, we propose a theoretical framework for the oligomerization kinetics of sequences at non-dilute conditions. This framework builds on non-equilibrium thermodynamics and can be used to explore high-dimensional sequence spaces.
Using our framework, we show that the oligomerization kinetics triggers the phase separation of sequences, which in turn creates a very strong bias for the enrichment and depletion of specific sequences.  A key finding is that, in sequence distributions at thermodynamic equilibrium, more cooperative sequences that interact well with their environment  are favored, while non-equilibrium fragmentation selects for low-information and less cooperative sequences.

Our findings can be tested experimentally considering nucleotide mixtures capable of DNA or RNA polymerization~\cite{mast_escalation_2013,rosenberger_self-assembly_2021,mutschler_freezethaw_2015}.  
For large enough concentrations of nucleotides, long enough sequences can form such that the mixing entropy is low enough to facilitate phase separation into oligomer-rich and oligomer-poor phases. 
Large enough salt concentrations screen  the negative charges of the DNA or RNA backbone, enabling sequence-specific, base-pair-mediated interactions at non-dilute conditions. 
Such interactions were shown to form liquid phases~\cite{aumiller_rna-based_2016,mitrea_phase_2016} and gels~\cite{sato_sequence-based_2020,xing_microrheology_2018}. 
Non-equilibrium fragmentation need not be engineered; for RNA, it is an omnipresent property on reasonable experimental time-scales of hours to days~\cite {goppel_thermodynamic_2022,calaca_serrao_high-fidelity_2024}.
Deep sequencing techniques  combined with HPLC  and mass spectroscopy can screen for more cooperative or less cooperative sequences using, for example, the cooperativity measure proposed in this work. 

Our theoretical work is relevant for the molecular Origin of Life~\cite{ianeselli_physical_2023,budin_expanding_2010}, and for designing \textit{de novo} life systems capable of undergoing Darwinian evolution~\cite{otto_approach_2022,kriebisch_roadmap_2024}. 
In particular, our finding of a sub-exponential effective occupation of the possible sequence space indicates a very strong selection pressure by simply considering non-equilibrium fragmentation. 
Our work shows that maintaining the system away from equilibrium sets the prerequisite for a rich exploration of the sequence space and, thus, Darwinian evolution on a molecular scale.
In particular, 
sequences of low cooperativity and low information are more likely to have exotic sequence patterns or motifs that may provide target-specific molecular functionalities, such as the catalysis of specific molecular components.

\section{Acknowledgements}

We are grateful for the insightful discussions with H.\ Vuijk, S.\ Gomez, M.\ Koul, and G.\ Granatelli.
We are also grateful for the discussions with 
A.\ Schmid,
F.\ Dänekamp,
S.\ Wunnava, 
A.\ Serrao, and  
P.\ Schwintek 
on how to experimentally scrutinize our theoretical predictions.
D.\ Braun and C.\ Weber thank the TRR 392: Molecular evolution in prebiotic environments
(Project number 521256690) for support. 
C.\ Weber acknowledges the 
European Research Council (ERC) for financial support under the European Union’s Horizon 2020 
research and innovation programme (``Fuelled Life'' with Grant agreement No.\ 949021).

\newpage
\appendix
\section*{Appendix}

\section{Most enriched mixed sequence with varying total volume fraction and fragmentation rate}\label{seq:len_dist}

\begin{figure}[tb]
    \centering
     \makebox[\textwidth][c]{\includegraphics[width=\textwidth]{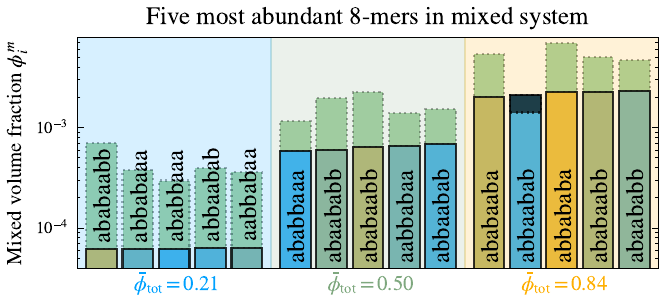}}
    \caption{\textbf{Phase separation enriches specific sequences and alters enrichment bias:} In a mixed system, the most abundant 8-mers homo-sub-strands of $aa$ and $bb$ appear more often than in the oligomer-rich phase, as displayed in Fig. \ref{fig:4}a. Phase separation generally increases the volume fractions (indicated by the dashed green bars), but for dense enough systems, depletion through phase separation is possible (indicated by the dashed black bar). Phase separation is not solely a magnifier of sequences but creates a bias.
     }
     \label{fig:8}
\end{figure}

Here, we discuss the difference between the phase-separated system and the mixed system for the five most abundant sequences when varying the total sequence volume fraction at chemical equilibrium. 
For the most abundant 8-mers in a mixed system, homo-sub-strands of $aa$ and $bb$ appear more often than in the oligomer-rich phase, as displayed in Fig. \ref{fig:4}a. As $\bar{\phi}_\text{tot}$ increases, the most enriched sequences change from more similar to the dilute phase (bluish) to more similar to the oligomer-rich phase (orangish). When $\bar{\phi}_\text{tot}$ approaches the oligomer-poor or -rich phase, the sequence distribution becomes the same. The mixed volume fractions of the most abundant sequences are generally much smaller than their corresponding quantity for the phase average (indicated by the dashed green bars). The relative increase through phase separation is the largest for smaller $\phi_\text{tot}$, as already observed in Fig. \ref{fig:4}c. With an increasing value of $\phi_\text{tot}$, the distributions become less flat, and a depletion through phase separation is possible (indicated by the dashed black bar). The fact that the most enriched sequences in the oligomer-rich phase differ from those enriched when mixed means that phase separation is not simply a magnifier of volume fraction but comes with a bias.

\begin{figure}
    \centering
    \makebox[\textwidth][c]{
    {\includegraphics[width=\textwidth]{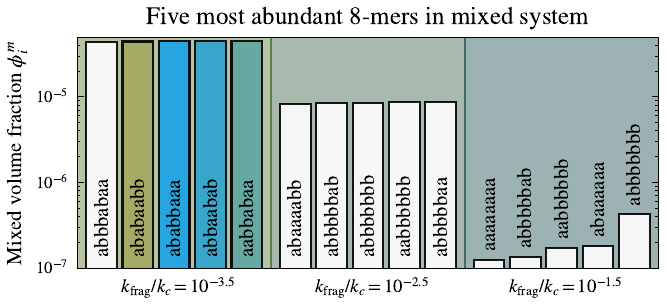}}}
     \caption{
     \textbf{Fragmentation enhances same sequences for mixed system, but less pronounced:} The most abundant $8$-mers behave similarly when mixed compared to when phase separated (displayed in Fig. \ref{fig:6}b), though a larger value of $k_\text{frag}$ is necessary. Still, the same trend of changing the selection bias from alternating sequences to longer blocks of the same monomer type with increasing $k_\text{frag}$ appears. 
     }
    \label{fig:9}
\end{figure}

For a mixed system, a similar sequence selection as for phase-separated systems (displayed in Fig.~\ref{fig:6}) occurs with increasing fragmentation rate, as seen in Fig.~\ref{fig:9}. For slow fragmentation rates, the most enriched sequences are similar to mixed systems at chemical equilibrium (displayed in Fig.~\ref{fig:4}b). With increasing $k_\text{frag}$ the most enriched 8-mers have longer blocks of the same monomer type; this happens concomitantly with the sequences distribution becoming flatter (Fig.~\ref{fig:6}d). For large fragmentation rates, the homopolymeric blocks almost span the size of the oligomer, as was observed with phase separation. For a mixed system, a larger value of $k_\text{frag}/k_c$ is necessary for the same type of sequence selection as for a phase-separated system.

\begin{figure}
    \centering
    \makebox[\textwidth][c]{
    {\includegraphics[width=\textwidth]{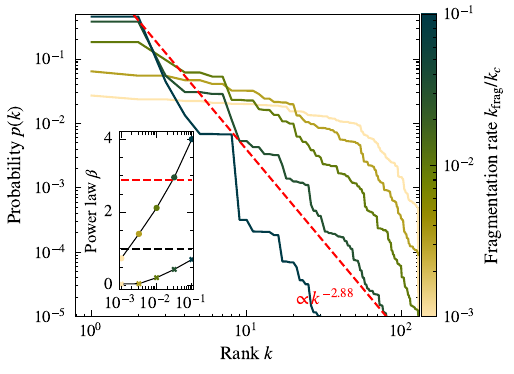}}}
     \caption{
     \textbf{The rank distribution transitions from an exponential to a power-law with fragmentation rate:} For a low fragmentation rate, the rank distribution of 8-mers follows an exponential. With large fragmentation rates, on the other hand, the distribution transitions to a power-law behavior $p(k)\propto k^{-\beta}$. The exponent $\beta$ increases strongly with the fragmentation rate $k_\text{frag}$ (inset) if the system is phase separated (upper curve). The example exponent of $2.88$ is displayed in the inset together with Zipf's law exponent of 1 \cite{sole_zipflaw2002}. For a mixed system (lower curve inset), $\beta=1$ is not reached.
  }
    \label{fig:10}
\end{figure}
To verify the transition from an exponential decay in the rank distribution, as displayed in Fig.~\ref{fig:6}d, to a power-law decay, the same data is plotted in a log-log plot in Fig.~\ref{fig:10}.
Once sufficiently far from equilibrium, the rank distribution approaches a straight line on a log-log representation, indicating a power-law decay. The best-fit exponent for the power law is displayed for different fragmentation rates in the inset, where the upper curve is phase-separated, and the lower is mixed. Note that the inset includes the best-fit power-law exponent for distributions with only a weak tendency of power-law. A power law exponent equal to one, as suggested by Zipf's law~\cite{sole_zipflaw2002}, is indicated as a horizontal line in the inset. 
 
\section{Thermodynamics of sequences: Free energy, interactions, and internal energy}\label{seq:F_chi_omega}

The chemical potentials $    {\mu}_{i} ( \{ \phi_j \}_{j\in\sigma}) = \nu_{i}\partial{f}/{\partial \phi_{i}}\vert_{\phi_{j\neq i}} 
  $ determine the oligomerization rates (Eq.~\eqref{eq:reaction_rate}) and are calculated from the free energy density $f$.
  We consider a free energy with a mixing entropy and interaction up to second order in the volume fraction $\phi_i$, often referred to as the Flory-Huggins free energy~\cite{bartolucci_interplay_2024,bauermann_chemical_2022,laha_chemical_2024}:
\begin{align}
    f &= \frac{k_BT}{\nu_s}\Bigg [ \sum_{i\in\sigma} \left\{\frac{\phi_{i}}{\rho_i}\log{\left(\frac{\phi_{i}}{\rho_i}\right)} + \frac{\omega_{i}\phi_{i}}{k_BT}\right\} + \frac{\omega_s\phi_s}{k_BT} \label{eq:free_energy}  \\ +& \phi_s\log{\left(\phi_s\right)}  + \frac{1}{2}\sum_{i\in\sigma}\sum_{j\in \sigma}\frac{\chi_{ij}\phi_i\phi_j}{k_BT} + \phi_s\sum_{i\in\sigma}\frac{\chi_{is}\phi_i}{k_BT} \Bigg ].  \nonumber
\end{align}
For two monomer types $a$ and $b$, the number of oligomers grows exponentially with the maximal length $L$; $\abs{\sigma}=2(2^L -1 )$. We assume the volumes of the two monomeric units to be equal, $\nu_a=\nu_b=\nu_1$, and the volume of oligomers to increase linearly with the number of constituting monomers $n_i$, as $\nu_i=\nu_1n_i$. Consequently, the relative oligomer molecular volumes $\rho_i =\nu_i/\nu_s$ also scale linearly with the number of monomers $n_i$. The first four terms in equation \eqref{eq:free_energy} represent the mixing entropy and internal energy per length $\omega_i$ of the oligomers and the solvent. Whereas the last two terms describe the mean field Flory-Huggins interactions between two oligomers $i$ and $j$ as $\chi_{ij}$, and between each oligomer and the solvent $s$ as $\chi_{is}$, relative to the thermal energy $k_BT$. As the sum over all volume fractions must add to one, the solvent volume fraction $\phi_s^\alpha$ is set by the total volume fractions $\phi_s^\alpha=1-\phi^\alpha_\text{tot}$.\par
\begin{figure}
    \centering
    \makebox[\textwidth][c]{
    {\includegraphics[width=\textwidth]{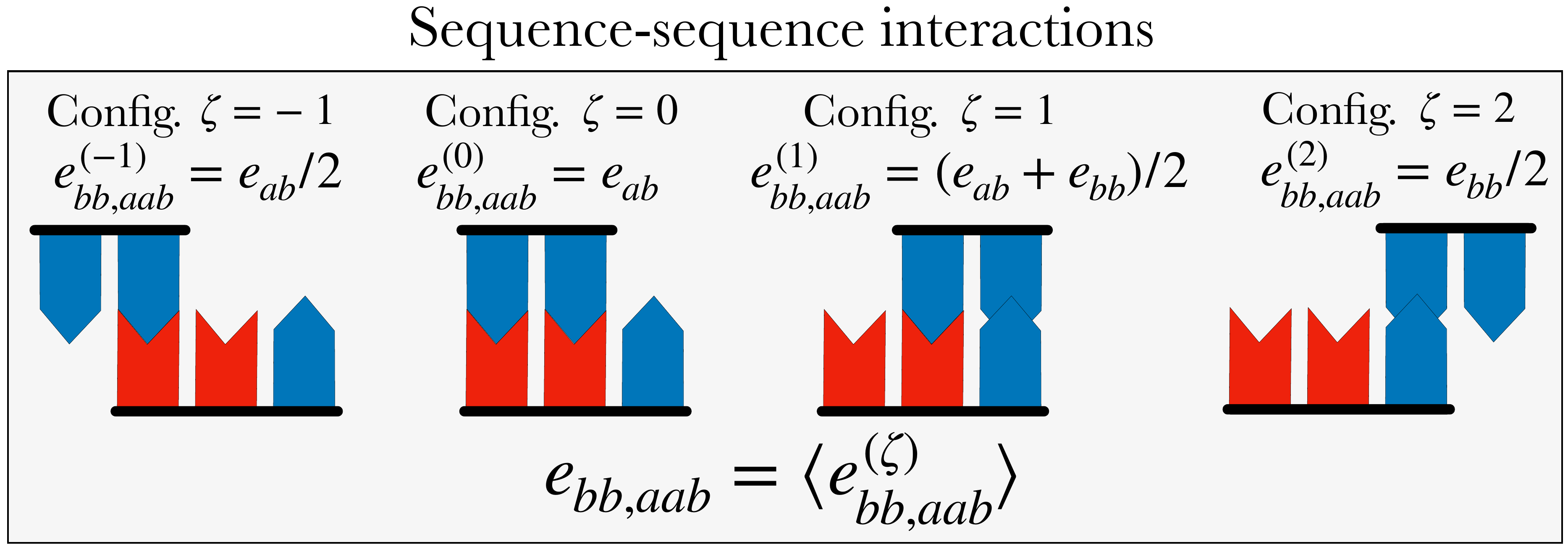}}}
    \caption{
     \textbf{Sequence interactions is calculated through a sliding average:} To calculate the interaction strength $e_{ij}$ between two sequences, here $i=bb$ and $j=aab$, we perform an average sliding configurations $\zeta$. Each sliding configuration has an associated interaction energy $e^{(\zeta)}_{ij}$ given by the total energy of the overlapping base-pairs interactions $(e_{aa},\,e_{ab}\,e_{bb})$ for that given configuration \eqref{eq:interact_case}. The average over all sliding configurations $\langle e_{ij}^{(\zeta)} \rangle$ is a Boltzmann average \eqref{eq:interact_boltzmann}, such that favorable configurations (here $\zeta=0$) contribute more to the average. Both orientations are considered, though not depicted here, due to the inversion symmetry of $bb$.
     }
    \label{fig:append_boltzmann}
\end{figure}

 
The Flory-Huggins interactions parameter between oligomers $i$ and $j$ is defined as $\chi_{ij} = e_{ij}-(e_{ii}+e_{jj})/2$ \cite{weber_physics_2019}, where $e_{ij}$ gives the effective interaction energy between the two oligomers. The value for $e_{ij}$ is determined by performing a Boltzmann average over all possible interaction energies $\langle e_{ij}^{(\zeta)} \rangle$ when sliding on oligomer along the other. The interaction strength for a specific configuration $\zeta$ can be written as (Fig.~\ref{fig:append_boltzmann}):
\begin{equation}
     e_{ij}^{\zeta,\pm} = \frac{1}{\underline{n_{ij}}}\sum\limits_{\beta=1}^{\underline{n_{ij}}}
    \begin{cases}
    0  \,\,\,\,\,\,\text{if $\,\,\Gamma_{i,\beta}=0$  or $\Gamma_{j,\beta\pm\zeta}=0$} \\
        e_{aa}\,\, \text{if $\,\,\Gamma_{i,\beta}=\Gamma_{j,\beta\pm\zeta}=a$} \\
        e_{ab}\,\, \text{if $\,\,\Gamma_{i,\beta}\neq\Gamma_{j,\beta\pm\zeta}\neq 0$} \\
        e_{bb}\,\, \text{if $\,\,\Gamma_{i,\beta}=\Gamma_{j,\beta\pm\zeta}=b$}
    \end{cases} \, ,\label{eq:interact_case}
\end{equation}
where $\Gamma_{i,\beta}$ picks the monomer-type at position $\beta$ of sequence $i$. If there is an overhang, and therefore no unit to pair up with, $\Gamma_{i,\beta}$ takes a value of $0$. To get the average interaction per length, the sum is normalized by the upper limit of the number of base-pair interactions, which is the length of the shortest of the two sequences, denoted $\underline{n_{ij}}=\text{min}(n_i, n_j)$. The sum thus adds up the total interaction energy for a given configuration, assuming pairwise  interactions; these interactions are similar to a Watson-Crick  for the case of DNA and RNA. The pairwise  interaction strengths $e_{aa}$, $e_{ab}$, and $e_{bb}$, are input parameters populating the full interaction matrix. This matrix can be calculated prior to the numerical studies for a given maximal length $L$, setting the number of possible sequences $|\sigma|$.  An illustration of the interaction energy of a specific configuration $k$ is displayed in Fig.~\ref{fig:1}. Averaging over all possible pair-wise configurations, the effective interaction energy is found from the Boltzmann average:
\begin{equation}
     e_{ij} = \frac{\sum\limits_{\pm}\,
     \sum\limits_{\zeta=-\underline{n_{ij}}+1}^{\overline{n_{ij}}-1}
     e_{ij}^{\zeta,\pm}\exp{-\frac{e_{ij}^{\zeta,\pm}}{k_BT}}
     }{\sum\limits_{\pm}\,\sum\limits_{\zeta=-\underline{n_{ij}}+1}^{\overline{n_{ij}}-1}
    \exp{-\frac{e_{ij}^{\zeta,\pm}}{k_BT}}}, \label{eq:interact_boltzmann}
\end{equation}
where the sum over $(\pm)$ takes both orientations into account, and the second sum is over the $(n_i+n_j-1)$  possible sliding configurations per orientation, where $\overline{n_{ij}}$ gives the largest length of the two sequences $\overline{n_{ij}}=\text{max}(n_i,n_j)$.\par 
The solvent interaction is assumed to be equal to the solvent interaction of its constituting monomeric units, $e_{is} = (a_ie_{as} + b_ie_{bs})/n_i$, entering the Flory-Huggins interactions parameter as $\chi_{is}=e_{is}-(e_{ii} + e_{ss})/2$, where $a_i$ and $b_i$ are the number of monomers of type $a$ and $b$ in sequence $i$.

A key contribution to the free energy \eqref{eq:free_energy} are the internal energies per length $\omega_i$. We assign to each monomer type  an intrinsic internal energy $\Omega_a$ and $\Omega_b$, and let each monomer in the oligomer interact with its neighbors through the back-bone bond energies $\epsilon_{aa}$, $\epsilon_{bb}$, $\epsilon_{ab}$, such that
\begin{align}
\begin{split}
    \omega_{i} &= \frac{a_i\Omega_a + b_i\Omega_b}{n_i} + \frac{e_{ii}}{2} \\ &+ \frac{n_{aa}^{(i)}\epsilon_{aa} + n_{bb}^{(i)}\epsilon_{bb}+n_{ab}^{(i)}\epsilon_{ab}}{n_i},
\end{split}
\end{align}
where $n_{jk}^{(i)}$ is the number of $(jk)$ neighbors along oligomer $i$. The self-interaction term $e_{ii}/2$ is included to retrieve an unbiased sequence distribution, i.e., flat, in the dilute limit, where $\bar{\phi}_s\rightarrow 1$. In this work, we treat all backbones equally (sequence-independent), such that $\epsilon_{aa}=\epsilon_{bb}=\epsilon_{ab}=\epsilon$. This way, the backbone has no preference towards specific sequence patterns, making all bias shown in this work exclusively arise from interactions among sequences. Decreasing the value of $\epsilon$ will result in a larger average length. More parameters for next neighbor correlation could be introduced to better model sequence distributions in peptides~\cite{toal_randomizing_2015,mu_effects_2007}.

\section{Oligomerization at chemical equilibrium}

\begin{figure*}[tb]
    \centering
    \makebox[\textwidth][c]{
   {\includegraphics[width=\textwidth]{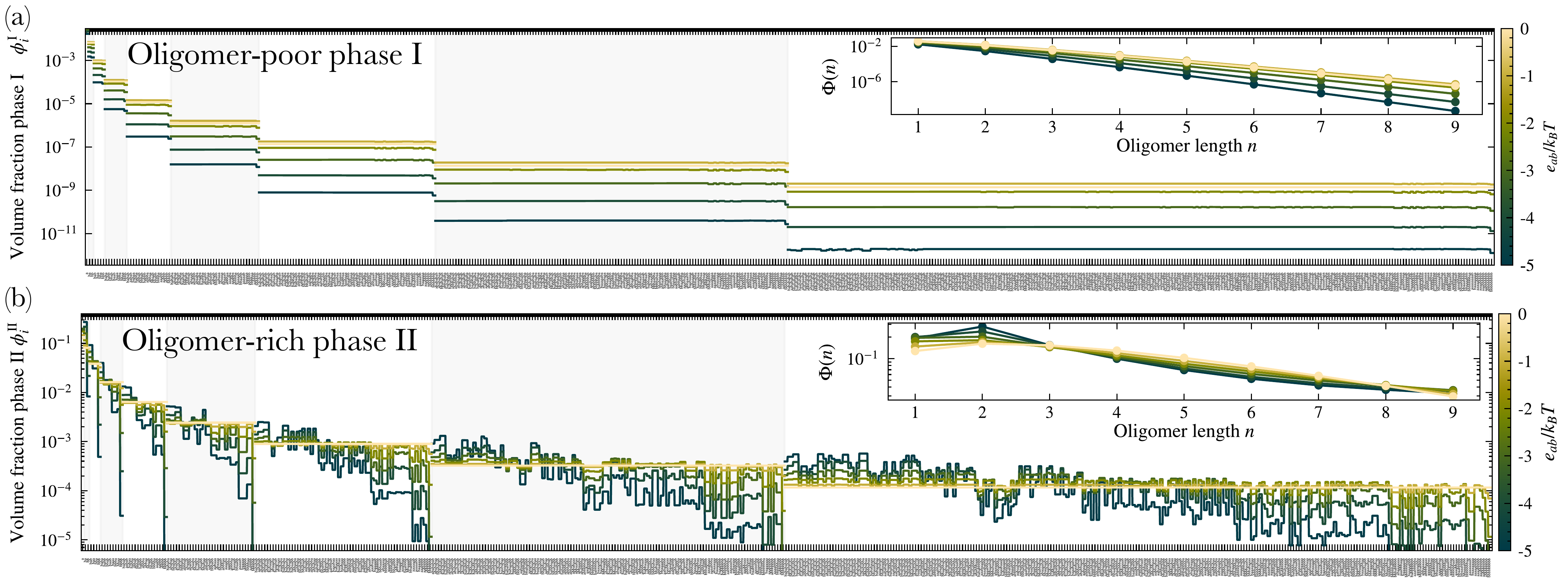}}}
     \caption{
     \textbf{
    Formation of alternating sequences requires phase separation and 
     strong enough sequence interactions. }
     Sequence distributions in the oligomer-poor phase I (a) and the oligomer-rich phase II (b). While the sequences are approximately equally distributed in the oligomer-poor phase I for each oligomer length $l$, there are strong variations in the oligomer-rich phase II. The prerequisite for the enrichment or depletion of specific sequences via phase separation is a strong enough interaction strength $e_{ab}$ (values in $k_BT$ indicated by color code).
     Sequences are ordered by the number of $a$-$b$-neighbors, and sequences of equal number of $a$-$b$-neighbors are sub-ordered by their Hamming distance to the perfectly alternating sequence of equal length.
     Inset: The length distribution of oligomers is weakly affected by the interaction strength.
    }
\label{fig:5_appendix}
\end{figure*}

At chemical equilibrium, the chemical potentials of all sequences $i$ are given by the chemical potential of the monomers \begin{equation}
     \mu_{i}^\alpha = a_i\mu_a^\alpha + b_i\mu_b^\alpha \, ,\label{eq:chem_eq}
\end{equation}
where $a_i$ is the number of monomers of type $a$ in oligomer $i$, and $b_i$ correspondingly for type $b$, such that $n_i=a_i+b_i$.
This condition represents that at chemical equilibrium, the chemical potential of oligomer $i$ must be equal to the chemical potential of its constituting monomers. When this condition is fulfilled, all oligomerization fluxes vanish ($r_i=0$). If this condition is concomitantly satisfied with the conditions of phase coexistence, the system is at thermodynamic equilibrium. Mixed systems can satisfy chemical equilibrium but will not be at thermodynamic equilibrium.\par

From the chemical equilibrium condition, we see that it is the chemical potential of the monomeric units that determine the chemical potential of all other sequences \eqref{eq:chem_eq}, and hence their volume fraction. This leaves two independent degrees of freedom, namely the total volume fraction for each nucleotide basis,
\begin{equation}
    \phi_A^\alpha = \sum_{i\in\sigma} \frac{a_i\phi_{i}^\alpha}{n_i},\quad\qquad
    \phi_B^\alpha = \sum_{i\in\sigma} \frac{b_i\phi_{i}^\alpha}{n_i}. \label{eq:def_conserved_quantities}
\end{equation}
These two quantities uniquely determine the chemical equilibrium sequence distribution in a given phase. As oligomerization conserves nucleotides, the total oligomer volume fraction of each nucleotide type, namely $\bar{\phi}_A$ and $\bar{\phi}_B$, are conserved during the kinetics and must be specified when initializing the system. To find the length distribution, the total volume fraction can be decomposed into a total oligomer volume fraction for each length
\begin{equation}
    \Phi^\alpha(l) = \sum_{i\in\sigma}\phi_i^\alpha\delta_{l,n_i} \label{eq:phi_tot_each_l} \,.
\end{equation}
To determine the sequence distribution, we must calculate the chemical potentials. From the free energy \eqref{eq:free_energy}, the chemical potential of an oligomer \eqref{eq:chem_eq} reads
\begin{align}
\begin{split}
    \frac{\mu_{i}}{\rho_i\,k_BT} &= \frac{1}{\rho_i}\log{\left(\frac{\phi_i}{\rho_i}\right)} - \log{\left(\phi_s\right)} + \frac{1}{\rho_i} - 1 \\ &+\frac{\omega_i-\omega_s}{k_BT}+ \sum_{j\in\sigma}\phi_j\frac{\chi_{ij} - \chi_{js}}{k_BT}  + \frac{\phi_s\chi_{is}}{k_BT} \, . \label{eq:chem_pot}
\end{split}
\end{align}
Using the chemical equilibrium condition \eqref{eq:chem_eq}, the sequence distribution is found to be
\begin{align}
\begin{split}
    &\phi_i^\alpha = \rho_i\left(\frac{\phi_a^\alpha}{\rho_1}\right)^{a_i}\left(\frac{\phi_b^\alpha}{\rho_1}\right)^{b_i} \exp{\rho_i - \rho_1} \\ \label{eq:equil_phi} &\times \exp{-\frac{\rho_i}{k_BT}\left(\omega_i-\frac{a_i\omega_a + b_i\omega_b}{a_i+b_i} \right) } \\
    &\times\exp{-\frac{\rho_i\phi_s^\alpha}{k_BT}\left(\chi_{is} - \frac{a_i\chi_{as} + b_i\chi_{bs}}{a_i+b_i}\right)} \\ &\times \exp{-\frac{\rho_i}{k_BT}\sum_{j\in\sigma}\phi_j^\alpha\left[\chi_{ij} - \frac{a_i\chi_{aj} + b_i\chi_{bj}}{a_i+b_i}\right]} \, .
    \end{split}
\end{align}
The first line represents the mixing entropy, resulting in an exponential decay with length. The remaining terms characterize the energy change from distributing $a_i$ and $b_i$ monomers into an oligomer $i$, changing the internal energy $\omega_i$, and interaction energies with the solvent $\chi_{is}$, and among sequences $\chi_{ij}$. The more negative the energy change, the larger $\phi_i^\alpha$ will be.

With our choice of interactions and internal energies, one can write the equilibrium distribution as 
\begin{align}
    \phi_i^{\alpha}  &= \rho_i\left(\frac{\phi_a^{\alpha}}{\rho_1}\right)^{a_i}\left(\frac{\phi_b^{\alpha}}{\rho_1}\right)^{b_i} \exp{\left(\rho_i-\rho_1\right)\left( 1-\frac{\epsilon}{k_BT} \right) }  \label{eq:equil_phi_es}
    \\ 
    & \times \exp{-\frac{\rho_i}{k_BT}\sum_{j\in\sigma}\phi_j^{\alpha} \left[e_{ij} - \frac{a_ie_{aj} + b_ie_{bj}}{a_i+b_i}\right]} \, .\nonumber
\end{align}
Notably, the solvent interactions $e_{is}$ do not affect the chemical equilibrium sequence distribution, though they affect the chemical reaction kinetics and phase compositions $\phi_A^\alpha$, $\phi_B^\alpha$. For two oligomers of equal length in a phase where $\phi^\alpha_A=\phi^\alpha_B$, or alternatively $a_i=a_j$ and $b_i=b_j$, we find that the non-solvent interaction terms solely set the relative volume fraction
\begin{align}
    &\frac{\phi_i^\alpha}{\phi_j^\alpha} = \exp{-\frac{\rho_i}{k_BT}\left(\sum_{k\in\sigma}\phi_k^\alpha\left[ e_{ik} - e_{jk} \right]  \right)} \label{eq:rel_prop} \\
 &\times \exp{-\frac{\rho_1}{k_BT}\sum_{k\in\sigma}\phi_k^\alpha\left( e_{ak}\left(a_i - a_j\right)  + e_{bk}\left(b_i - b_j\right) \right) \nonumber
}\, .  
\end{align}
We see that in a dilute system where $\bar{\phi}_\text{tot}\rightarrow 0$, all sequences of each length will be equally abundant. As the total sequence volume fraction $\bar{\phi}_\text{tot}$ increases, the oligomer-oligomer interactions can co-enrich sequences that adapt to the sequence patterns in the respective phase, leading to more cooperative sequences. Each sequence is affected through interactions with all other sequences and likewise affects them in return. \par 

The full sequence distribution in the oligomer poor and oligomer rich phases at thermodynamic equilibrium is depicted in Fig. \ref{fig:5_appendix}. As Eq. \eqref{eq:rel_prop} predicts, the oligomer poor phase (\ref{fig:5_appendix}a) does not contain sufficient oligomer material in order for enrichment or depletion of specific sequences. Thus, entropy dominates, leading to sequences of the same length being equally distributed, independent of the interaction strength $e_{ab}$. This is not the case in the oligomer-rich phase (Fig. \ref{fig:5_appendix}b), where a strong enrichment and depletion is observed. For each length, the $x$-axis is ordered after the number of $a$-$b$-neighbours. The sequence distribution shows the general trend of an enrichment of sequence with many $a$-$b$-neighbours, and depletion of those with few. 

\par 

Equation \eqref{eq:rel_prop} can further be used as a design principle for $\chi_{ij}$ to realize a specific sequence distribution of $n$-mers given some sequence distribution for $n_i<n$. Since monomers and shorter sequences have the largest volume fractions, they will dominate the contribution for determining $\phi_i^\alpha/\phi_j^\alpha$. Furthermore, since longer sequences are generally dilute, their effect on the volume fraction of shorter sequences can be disregarded. Thus, one might tune $\chi_{ij}$ between the shorter sequences and the $n$-mers for which a specific sequence distribution is desired by only summing for sequences where $n_k<n$ in Eq. \eqref{eq:rel_prop}. 
\par 
The chemical potentials of each sequence can be expressed in terms of the reference chemical potential $\mu_i^0$, which is composition independent, and the activity coefficient $\gamma_i$, containing the contributions from the interactions among sequences:
\begin{equation}
    \mu_i = \mu_i^0(T, V) + k_BT\log{\left(\phi_i\, \gamma_i\{\{\phi_i\}, T, V\}\right)} \, .
\end{equation}
Through these quantities, one can define a cooperativity $\Lambda_i$, measuring how well adapted a specific sequence is to its environment, as defined in Eq. \eqref{eq:fitness}. Using the expression for the chemical potential \eqref{eq:chem_pot}, one finds that
\begin{align}
    \mu_i^0 &= 1-\rho_i-\log{\rho_i}+\rho_i\frac{\omega_i-\omega_s}{k_BT}, \label{eq:chem_pot_decomp}\\
    \gamma_i^\alpha &= \left(\frac{1}{\phi_s^\alpha}\right)^{\rho_i}\exp{\rho_i\frac{\sum_{j\in\sigma}\phi_j^\alpha\left( \chi_{ij}-\chi_{js} \right) + \phi_s^\alpha\chi_{is}}{k_BT}} \, .\nonumber
\end{align}
The ratio of the activity coefficients of each phase  sets the partitioning coefficient, 
$\phi_i^\RN{1}/\phi_i^\RN{2}=\gamma_i^\RN{2}/\gamma_i^\RN{1}$~\cite{bauermann_chemical_2022}. At thermodynamic equilibrium, it is the component with the largest cooperativity that becomes the most abundant.

\section{Parameters and validity of theory}\label{seq:limits}

We consider the case where phase equilibrium holds at each time during the chemical kinetics.
As a result, each thermodynamic quantity is homogeneous in each phase. This simplified case is satisfied if 
diffusion rates of reacting components are fast compared to the time scales of the chemical rates and fragmentation rate coefficients:
\begin{equation}
    k_c,\,\,k_\text{frag} \ll \frac{D_{\text{min}}}{l^2} \, , \label{eq:asump}
\end{equation}
where $l$ is the characteristic system size and $D_{\text{min}}$ is the smallest diffusion coefficient of all the components.

The validity of this approximation must be understood through kinetic measurements of diffusion coefficients and chemical reaction rates. For a diffusion coefficient of $100\,\mu\text{m}^2/\text{s}$ and a reaction time scale of minutes, this is satisfied for system sizes of $0.1$mm. The parameters used are listed in table \ref{tab:params}.

\begin{table*}[]
 \caption{\textbf{The interaction strengths $e_{ij}$ and $e_{is}$, internal energies $\epsilon$, $\Omega_i$, and $\bar{\phi}_A$ and $\bar{\phi}_B$, used for all figures.} Common for all figures in this work is that $\rho_1=1$, $k_c=1$, $\omega_S=0$, and $k_BT=1$. All energies in this table are measured in units of $k_BT$. \label{tab:params}}
\begin{tabular}{@{}lllllllllllll@{}}
\toprule
Figure $\quad$& $e_{AA}\quad$ & $e_{BB}\quad$ & $e_{AB}\qquad$ & $e_{AS}\qquad$ & $e_{BS}\qquad$  & $e_{SS}\qquad$ & $\epsilon \qquad$ & $\Omega_{A}\qquad$ & $\Omega_{B}\qquad$ & $\bar{\phi}_A\qquad$ & $\bar{\phi}_B\qquad$ & $L\qquad$\\ \midrule
\ref{fig:3}     &    0      &    0      &   -5           &  -0.92      &  -0.92   & -3.35 & -1.0  & 2 & 2 & 0.03 & 0.03 & 5 \\
\ref{fig:4}     &    0      &    0      &   -5           &  -0.92      &  -0.92   & -3.35 & -1.0  & 1 & 1 & -- & -- & 8 \\
\ref{fig:5}     &    0      &    0      &   $e_{ab}$     &  $2.005 + e_{ab}/4$    &  $2.005 + e_{ab}/4$   & -3.35 & -0.6  & 0.6 & 0.6 & -- & -- & 9 \\
\ref{fig:5.5}     &    0      &    0      &   -5           &  -0.92      &  -0.92   & -3.35 & -1.0  & 1.0 & 1.0 & 0.11 & 0.11 & 8 \\
\ref{fig:6}     &    0      &    0      &   -5           &  -0.92      &  -0.92   & -3.35 & -1.0  & 1.0 & 1.0 & 0.11 & 0.11 & 8 \\
\ref{fig:7}     &    0      &    0      &   -5           &  -0.92      &  -0.92   & -3.35 & -1.0  & 1.0 & 1.0 & 0.11 & 0.11 & 8 \\
\ref{fig:8}     &    0      &    0      &   -5           &  -0.92      &  -0.92   & -3.35 & -1.0  & 1.0 & 1.0 & 0.11 & 0.11 & 8 \\
\ref{fig:9}     &    0      &    0      &   -5           &  -0.92      &  -0.92   & -3.35 & -1.0  & 1.0 & 1.0 & -- & -- & 8 \\
\ref{fig:10}     &    0      &    0      &   -5           &  -0.92      &  -0.92   & -3.35 & -1.0  & 1.0 & 1.0 & 0.11 & 0.11 & 8 \\
\ref{fig:5_appendix}     &    0      &    0      &   $e_{ab}$     &  $2.005 + e_{ab}/4$    &  $2.005 + e_{ab}/4$   & -3.35 & -0.6  & 0.6 & 0.6 & -- & -- & 9 \\
\bottomrule
\end{tabular}
\end{table*}

\section{Oligomerization kinetics at phase equilibrium}\label{sec:appendix_BLkinetics}

Using the method derived by Bauermann et al. \cite{bauermann_chemical_2022}, one can find the diffusive flux $j_i^\alpha$ between the phases such that the system remains at phase coexistence. Here, we summarize their key equations and apply them to our reaction network. For a full description, see their complete work \cite{bauermann_chemical_2022}.\par 
In this section, we derive the expression for the diffusive flux $j_i^\alpha$ between the phases. To find this flux, we consider  systems where the time scale of each sequence to diffuse the system size $l$ with a diffusion coefficient $D$, $l^2/D$, is short compared to  the oligomerization time scale, $k_c^{-1}$, meaning $ l \ll \sqrt{D/k_{c}}$. For phase-separated systems, this implies that the volume fraction profiles of each sequence in both phases are flat. Since the system must satisfy the conditions at all times, the oligomerization kinetics is confined to the binodal manifold (Fig.~\ref{fig:1_react_network}(c)).

For two-phase coexistence, this means that the chemical potentials in the two phases must follow
\begin{equation}
    \partial_t \mu_i^\RN{1} = \partial_t \mu_i^\RN{2}, \label{eq:equal_chem}
\end{equation}
where $i\in\sigma$ contains monomers and sequences but not the solvent. Each side of the equation above can be rewritten using that the chemical potential of component $i$ depends on the composition of all other components, ${\mu}_{i} ( \{ \phi_j \}_{j\in\sigma})$:
\begin{equation}
     \partial_t \mu_i^\alpha = \sum_{k\in\sigma} \dot{\phi_k^\alpha}\pdv{\mu_i^\alpha}{\phi_k} \, .
\end{equation}
The time derivative of the volume fraction for incompressible systems, where the molecular volumes are constant in time $\text{d}_t\nu_i=0$, follow Eq.~\eqref{eq:kinetic_eq}. Using the expression for the volume change of the phase (Eq.~\eqref{eq:vol_change}), one obtains
\begin{equation}
    \dot{\phi_k^\alpha} = r_k^{\alpha} - j_k^{\alpha} + \phi_k^{\alpha} \sum_{l=s,\sigma} j_l^{\alpha} \, . \label{eq:kinetic_in_app}
\end{equation}
Using this expression, we write the time-derivative of the chemical potentials as
\begin{equation}
     \partial_t \mu_i^\alpha = \sum_{k\in\sigma} \left( r_k^{\alpha} - j_k^{\alpha}  + \phi_k^{\alpha} \sum_{l\in s,\sigma} j_l^{\alpha} \right)\pdv{\mu_i^\alpha}{\phi_k}\, ,
\end{equation}
that can be expressed in terms of the fluxes: 
\begin{align}
    \begin{split}
    \partial_t \mu_i^\alpha &= \sum_{k\in\sigma} j_k^\alpha\left(-\pdv{\mu_i^\alpha}{\phi_k} + \sum_{l\in\sigma} \phi_l^\alpha \pdv{\mu_i^\alpha}{\phi_l}\right) \\ &+ j_s\sum_{k\in\sigma} \phi_k\pdv{\mu_i^\alpha}{\phi_k} + \sum_{k\in\sigma} \pdv{\mu_i^\alpha}{\phi_k}r_k^\alpha \, .\end{split}
\end{align}
Using the kinetic constraint of the chemical potentials (Eq.~\eqref{eq:equal_chem}), we find: 
\begin{align}
     &\sum_{k\in\sigma} j_k^\RN{1}\left\{ \sum_{l\in\sigma} \phi_l^\RN{1} \pdv{\mu_i^\RN{1}}{\phi_l^\RN{1}} -\pdv{\mu_i^\RN{2}}{\phi_k^\RN{2}} + \frac{V^\RN{1}}{V^\RN{2}}\left(\sum_{l\in\sigma} \phi_l^\RN{2} \pdv{\mu_i^\RN{2}}{\phi_l^\RN{2}} -\pdv{\mu_i^\RN{2}}{\phi_k^\RN{2}} \right) \right\} \nonumber \\ &+ j_s^\RN{1}\left(\sum_{k\in\sigma} \phi_k^\RN{1}\pdv{\mu_i^\RN{1}}{\phi_k^\RN{1}} + \frac{V^\RN{1}}{V^\RN{2}}\sum_{k\in\sigma} \phi_k^\RN{2}\pdv{\mu_i^\RN{2}}{\phi_k^\RN{2}}\right) \label{eq:chem_pot_constraint} \\ &= -\sum_{k\in\sigma} \pdv{\mu_i^\RN{1}}{\phi_k^\RN{1}}r_k^\RN{1} +  \sum_{k\in\sigma} \pdv{\mu_i^\RN{2}}{\phi_k^\RN{2}}r_k^\RN{2} \, . \nonumber
\end{align}
To obtain the equation above,  we have used conservation of particle number of the diffusive flux, implying
\begin{equation}
    j_i^\RN{1}V^\RN{1} = -j_i^\RN{2}V^\RN{2} \, . \label{eq:particle_cons}
\end{equation}
Eq.~\eqref{eq:chem_pot_constraint} is satisfied for all $i\in\sigma$, leaving one unknown, namely the diffusive flux of the solvent. It is determined  by the balance of the osmotic pressure,
\begin{equation}
    \Pi^\alpha = f(\{\phi_i^\alpha\}) - \sum_{i\in\sigma} \frac{\phi_i\mu_i}{\nu_i}. \label{eq:osmotic_pressure_def}    
\end{equation}
with a time evolution in each phase $\alpha=\text{I}, \text{II}$ given as:
\begin{equation}
    \partial_t \Pi^\alpha = -\sum_{k\in\sigma} \dot{\phi_k^\alpha}\theta_k^\alpha,\qquad\quad \theta_k^\alpha= \sum_{l\in\sigma} \frac{\phi_l^\alpha}{\nu_l}\pdv{\mu_l^\alpha}{\phi_k^\alpha} \, .
\end{equation}
Using the time-derivative of the volume fractions (Eq.~\eqref{eq:kinetic_in_app}), we find
\begin{align}
\begin{split}
    \partial_t \Pi^\alpha =& \sum_{k\in\sigma} j_k^\alpha\left(\theta_k - \sum_{l\in\sigma}\phi_l^\alpha\theta_l^\alpha\right) \\ &- j_s^\alpha\sum_{l\in\sigma}\phi_l^\alpha\theta_l^\alpha - \sum_{l\in\sigma}\theta_l^\alpha r_l^\alpha \, .
\end{split}
\end{align}
Equating the time derivative of the osmotic pressure in each phase $\partial_t\Pi^\RN{1}=\partial_t\Pi^\RN{2}$, leaves us with 
\begin{align}
    &\sum_{k\in\sigma} j_k^\RN{1}\left\{ \sum_{l\in\sigma}\phi_l^\RN{1}\theta_l^\RN{1} -\theta_k^\RN{1} + \frac{V^\RN{1}}{V^\RN{2}}\left( \sum_{l\in\sigma}\phi_l^\RN{2}\theta_l^\RN{2}- \theta_k^\RN{2}\right)  \right\} \label{eq:osmot_constraint} \\ & +j_s^\RN{1}\left( \sum_{l\in\sigma}\phi_l^\RN{1}\theta_l^\RN{1} + \frac{V^\RN{1}}{V^\RN{2}}\sum_{l\in\sigma}\phi_l^\RN{2}\theta_l^\RN{2}\right)   = \sum_{l\in\sigma}\left(\theta_l^\RN{2} r_l^\RN{2}-\theta_l^\RN{1} r_l^\RN{1}\right) \, .\nonumber
\end{align}
The set of equations in \eqref{eq:osmot_constraint} and \eqref{eq:chem_pot_constraint} can be written as a matrix-vector equation
\begin{equation}
    A_{ik}j^\RN{1}_k = s_i,
\end{equation}
which can be solved for $j^\RN{1}_k$. The reaction rates in each phase act as source terms in $s_i$, while $A_{ik}$ specifies how the composition changes in each phase will affect the chemical potentials and the osmotic pressure, constraining the diffusive fluxes between the phases to equalize the change in all quantities in each phase. 

\section{Net oligomerization rate with non-equilibrium fragmentation}\label{sec:app_NEQ}

The net rate $\tilde{r}_i^\alpha$ contains the monomer-pickup/release rate $r_i$ (Eq.~\eqref{eq:overall_reaction_rate}) that fulfills detailed balance of the rates (Eq.~\eqref{eq:det_balance}), and the non-equilibrium fragmentation rate $h^\alpha_{i\rightharpoonup j+\abs{i-j}} $ (Eq.~\eqref{eq:hydro_cut}): 
\begin{align}
    \tilde{r}_i^\alpha = r_i^\alpha - &\sum_{j\in\sigma_i} h^\alpha_{i\rightharpoonup j+\abs{i-j}}  \label{eq:total_react} \\ + &\sum_{j\in\sigma_o} \frac{n_i}{n_j} h_{j\rightharpoonup i+\abs{i-j}}^\alpha (1+\delta_{i,\abs{i-j}})\delta_{i\in\sigma_j} \, , \nonumber
\end{align}
where $\sigma_i$ is the $(n_i-1)$ sub-oligomers of sequence $i$ resulting from a single fragmentation event. 
In the last term of the equation above, $\delta_{i\in\sigma_j}$, takes a value of $1$ if $i$ is a member of $\sigma_j$, and zero otherwise, and $\delta_{i,\abs{i-j}}$ is the Kronecker-delta.\par 
One can find and approximate expression  \eqref{eq:approx_NESS_dist} for the sequence distribution of long sequences at the non-equilibrium steady state (NESS) by using that all time derivatives of Eq. \eqref{eq:kinetic_eq} is zero. This results in the condition that
\begin{equation}
     \tilde{r}_i^\alpha - j_i^\alpha = 0
\end{equation}
concomitantly in both phases. Multiplying this condition by the respect phase volume yields, by utilizing Eq. \eqref{eq:particle_cons}, the simpler condition of 
\begin{equation}
   0 = \tilde{r}_i^\RN{1}V^\RN{1} + \tilde{r}_i^\RN{2}V^\RN{2} \, . \label{eq:SS_cond}
\end{equation}
For olgiomers of the maximal length $L$, the gain term from fragmentation in Eq. \eqref{eq:total_react} is zero, such that Eq. \eqref{eq:SS_cond} becomes
\begin{equation}
    0 = V^\RN{1}\left( r_i - \sum_{j\in\sigma_i} h^\RN{1}_{i\rightharpoonup j+\abs{i-j}}  \right) + V^\RN{2}\left( r_i - \sum_{j\in\sigma_i} h^\RN{2}_{i\rightharpoonup j+\abs{i-j}}  \right), \label{eq:SS_cond2}
\end{equation}
where the phase condition of the oligomerization rate has been omitted due to phase coexistence. From the diluteness of long sequences in the oligomer-poor phase (II), approximately no fragmentation of these sequences is occurring, such that
\begin{equation}
    \frac{V}{V^\RN{2}}r_i = \sum_{j\in\sigma_i} h^\RN{2}_{i\rightharpoonup j+\abs{i-j}} \, .  \label{eq:SS_cond3}
\end{equation}
Using the expression for $h_{i\rightharpoonup j+\abs{i-j}}^\alpha$ in Eq. \eqref{eq:hydro_cut}, and the definition of the reaction rate $r_i$ \eqref{eq:reaction_rate}, where the reverse pathway has been omitted,
\begin{equation}
    \frac{Vk_c}{V^\RN{2}}\sum_{j\in\mathcal{R}_i}\exp{\frac{\mu_j + \mu_m}{k_BT}} = k_\text{frag}\phi_s^\RN{2}\phi_i^\RN{2}\, .
 \label{eq:SS_cond4}
\end{equation}
Here, $\mathcal{R}_i$ is the set of sequences that can oligomerize into sequence $i$. Solving for $\phi_i^\RN{2}$ yields the final expression of Eq. \eqref{eq:approx_NESS_dist}.

\bibliography{SEQ.bib}{}
\bibliographystyle{unsrtnat}

\end{document}